\documentclass[aps,prb,superscriptaddress,twocolumn,longbibliography]{revtex4-2}

\usepackage[pdftex]{graphicx}
\usepackage{dcolumn}
\usepackage{bm}
\usepackage{color}
\usepackage{amsmath}
\usepackage{centernot}
\usepackage{nicefrac}
\usepackage{amssymb} 

\usepackage{enumitem}
\usepackage{dsfont}
\usepackage{cancel}
	
\newcommand{\ff}[1]{{\boldsymbol #1}}
\newcommand{\ca}[1]{{\cal #1}}
\newcommand{\bi}{\begin{itemize}}
\newcommand{\ei}{\end{itemize}}
\newcommand{\be}{\begin{equation}}
\newcommand{\ee}{\end{equation}}
\newcommand{\ba}{\begin{eqnarray}} 
\newcommand{\ea}{\end{eqnarray}}

\newcommand{\sgn}{\mathrm{sgn}}

\newcommand{\ket}[1]{\left\vert#1\right\rangle}
\newcommand{\bra}[1]{\left\langle#1\right\vert}
\newcommand{\bracket}[3]{\left\langle#1\middle\vert#2\middle\vert#3\right\rangle}

\usepackage[pdftex,colorlinks=true,linkcolor=blue,citecolor=blue,filecolor=blue]{hyperref}

\begin{document} 
  
\title{Kondo Screening and Indirect Magnetic Exchange through a Conventional Superconductor Studied by the Density-Matrix Renormalization Group}

\author{Cassian Plorin} 

\affiliation{I. Institute of Theoretical Physics, Department of Physics, University of Hamburg, Notkestra{\ss}e 9-11, 22607 Hamburg, Germany}

\affiliation{The Hamburg Centre for Ultrafast Imaging, Luruper Chaussee 149, 22761 Hamburg, Germany}

\author{Michael Potthoff}

\affiliation{I. Institute of Theoretical Physics, Department of Physics, University of Hamburg, Notkestra{\ss}e 9-11, 22607 Hamburg, Germany}

\affiliation{The Hamburg Centre for Ultrafast Imaging, Luruper Chaussee 149, 22761 Hamburg, Germany}

\begin{abstract}
The competition between the Kondo screening and indirect magnetic exchange in systems with two magnetic impurities coupled to a conventional s-wave superconductor gives rise to a nontrivial ground-state phase diagram. 
Here, we utilize the density-matrix renormalization group (DMRG) method and exploit the non-abelian spin-SU(2) symmetry to study the phase diagram for two quantum-spin-$\frac12$ impurities locally exchange coupled to large one-dimensional chains.
The nonlocal inter-impurity exchange is treated as an emergent Ruderman-Kittel-Kasuya-Yosida (RKKY) coupling.
We find qualitatively different phase diagrams for impurity spins coupled to sites with odd or even distances $d$ on the chain and a partial-Kondo-screened spin-doublet phase that extends over the whole range of local exchange couplings $J$ in the limit of weak superconducting pairing strength $\Delta$.
Our numerical studies are complemented by exact diagonalization of small (quantum-box) systems and by perturbative-in-$J$ computations of the $d$ and $\Delta$ dependent RKKY interaction.
It is thereby demonstrated that the specific system geometry is essential for our understanding of magnetic impurity interactions in superconducting hosts, and thus for insights into the control of quantum-state properties in nanoparticle systems and topological superconductivity.
\end{abstract} 

\maketitle

\section{Introduction}  

Unconventional properties of physical systems are often a consequence of competing mechanisms at comparable energy scales. 
The Doniach diagram \cite{Don77} represents a prime example. 
Here, the individual Kondo screening \cite{Kon64,Hew93} of magnetic moments locally exchange-coupled to a Fermi sea competes with their mutual nonlocal magnetic interaction, which is mediated by the same Fermi sea.
This Ruderman-Kittel-Kasuya-Yosida (RKKY) indirect exchange \cite{RK54,Kas56,Yos57} $J_{\rm RKKY} \propto J^{2}$ is accessible by perturbation theory in the strength of the local exchange coupling $J$, while the Kondo effect is intrinsically non-perturbative with the Kondo temperature $T_{\rm K} \propto e^{-1/J}$ as the characteristic energy scale. 
The resulting Doniach diagram is a classical topic \cite{JKW81,JV87,JVW88} of many-body theory and material science that has stimulated a continuous discussion and methodical advancements over several decades
\cite{FHS87,SSK90,SSK93,VBH02,ZCSV06,JGL11,TSRP12,MSL12,STP13,ABMF14,YY18,AF19,ZHL20}, see, e.g., Refs.\ \cite{ELA18,EA20} for an overview.

The Kondo-vs.-RKKY competition is expected to be fundamentally different, if the electronic structure of the host material is gapped.
For a single magnetic impurity in a finite system, i.e., for a ``Kondo box'' \cite{TKvD99}, and when the finite-size gap $\delta$ becomes comparable to the bulk Kondo temperature, the logarithmic Kondo correlations are truncated by the system size.
Hence, for $\delta \gtrsim T^{\rm (bulk)}_{\rm K}$, the impurity spin cannot be Kondo screened because there are no low-energy excitations with energies below $T^{\rm (bulk)}_{\rm K}$ available.
The physics resulting from the competition between $T_{\rm K}$ and $\delta$ has been studied extensively \cite{Sch01,Sch02,SA02,CB02,FRF03,RM06,HKM06,BBH10}.

For several magnetic impurities \cite{LVK05}, i.e., for a finite ``Kondo-vs.-RKKY box'', the truncated Kondo effect competes with an RKKY-like nonlocal exchange with an exponential distance dependence at large distances, reminiscent of the Bloembergen-Rowland-type \cite{BR55}. 
Particularly interesting is the ``on resonance'' case, where a singly occupied conduction-electron eigenstate right at the Fermi edge gives rise to a two-fold degenerate many-electron ground state and is thus available for the screening.
It has been shown that this results in an unconventional reentrant competition at weak $J$ and small $\delta$ \cite{SGP12}.
Later, this work has been generalized to an arbitrary number of magnetic moments in a metallic nanostructure \cite{SHPM15} and to the strong-$J$ limit \cite{SHP14}, where the low-energy physics is captured by generalized central-spin models. 
Ultrafast dynamical manipulations of the interplay between Kondo and RKKY exchange have been studied for small nanoring geometries \cite{YHV18}. 

Several quantum spins coupled to a conventional s-wave superconductor represents another highly interesting class of systems with a competition between Kondo screening and indirect RKKY exchange. 
The competition is controlled by a gapped electronic structure and turns into a competition of a truncated ``Kondo'' effect (with exponentially decaying Kondo correlations) with an attenuated indirect ``RKKY'' magnetic exchange, which is exponentially suppressed at large distances. 
Within standard BCS theory, the gap size $2\Delta$ is related to the pairing strength $\Delta$ and is thus easily controlled via its temperature dependence.
Opposed to Kondo-box systems, however, the ground state is a total-spin singlet at $J=0$, and thus there is no ``on resonance'' case. 
The ground-state properties are rather dominated by level crossings of subgap bound states \cite{BVZ06}.

Currently, there is a strong interest in chains of magnetic adatoms on superconductors \cite{SBS+20,KCZ+21,vOF21,FBBO21,DHLK21,MCL21,LRR+22,MGDW22,SMFvO22,ZZN+23}. 
The induced Yu-Shiba-Rusinov (YSR) bound states \cite{Yu65,Shi68,Rus69} hybridize to form a subgap energy band that can host Majorana zero modes. 
As a platform for Majorana zero modes \cite{NDBY13,BS13,PGvO13,KSYL13} and topological superconductivity \cite{PHK+19,FvOS21} such systems have attracted much attention recently. 
The simplest system in this context consists of just two magnetic adatoms in proximity to a conventional superconductor
\cite{RHP+18,CFH+18,KDOL18,BSR+21}.
The importance of treating magnetic impurities or adatoms as quantum spins for the construction of the superconducting extension of Doniach's phase diagram has been emphasized recently \cite{SSFvO22}. 

In this paper, we study the case of two quantum spins with $s=\frac12$ with antiferromagnetic exchange coupling $J$ to a conventional (BCS) superconductor with gap $2\Delta$.
The two-impurity Kondo model poses a highly nontrivial many-body problem, which somewhat simplifies in case of a superconducting host, because the Kondo scale is cut by $\Delta$ and since the gap leads to exponentially decaying correlations.
In the weak-$J$ regime the problem thus becomes accessible to standard perturbation theory. 
Still, an exact treatment requires state-of-the-art numerical methods, such as the numerical renormalization group (NRG) \cite{Wil75,KWW80,JV87,BCP08}. 
The generic zero-temperature phase diagram has been studied using NRG \cite{ZLL+10,ZBP11,YMW+14} and has served as an important orientation for subsequent investigations \cite{RHP+18,FBBO21,KCZ+21,vOF21,MGDW22,SMFvO22,ZZN+23}.
In particular, in a certain parameter regime, this exhibits a total-spin-$\frac12$ ``molecular doublet'' phase, which is delocalized between the magnetic impurities, and other phases with dominating Kondo or RKKY correlations. 
However, within NRG, for technical reasons and related to the assumption of a flat density of states (see Ref.\ \cite{YMW+14}, supplemental material), the inter-impurity indirect magnetic (RKKY) exchange $J_{\rm RKKY}$ is treated as an {\em independent} parameter. 

Here, we employ the density-matrix renormalization group (DMRG) \cite{Whi92,Sch11} to study the ground-state phase diagram. 
The inter-impurity coupling is treated as an {\em emergent} interaction that depends on the details of the model and, in particular, on the distance $d$ between the impurity spins. 
On the other hand, the DMRG phase diagram is less generic and will depend on the details of the specific system geometry. 
The natural choice in the DMRG context is that of a one-dimensional chain geometry, and this is adopted here. 
Although this implies that the NRG and the DMRG results cannot be quantitatively mapped onto each other, it will be important to check, whether or not the general phase diagram topologies are the same.
Due to the expected exponential decay of RKKY correlations at large distances, we focus on geometries with spin impurities locally exchange coupled to nearest and to next-nearest neighbor sites of the chain. 

Our DMRG study is supplemented by and compared with analytical perturbative-in-$J$ computations of the RKKY interaction for weak $J$.
The computed RKKY exchange exhibits an unconventional distance dependence, i.e., a crossover from oscillatory behavior to exponential decay as function of $d$. 
Furthermore, the dependence on the pairing strength $\Delta$ turns out nontrivial with a sign change as function of $\Delta$ for even $d$.

It is common to treat the pairing strength as an adjustable parameter. 
Within BCS mean-field theory, this can be justified since there is a one-to-one correspondence to the strength of the effective attractive electron-electron interaction. 
The approximation must be questioned, however, if impurities are present, since this leads to a site-dependent pairing strength $\Delta_{i}$. 
On a qualitative level this issue is studied for a model with $L=3$ sites only, but where a site-dependent pairing strength $\Delta_{i}$ can be determined in a self-consistent way.

The paper is organized as follows:
Section \ref{sec:mod} introduces the model and the relevant symmetries. 
Some details of the DMRG are discussed in Sec.\ \ref{sec:dmrg}.
In Sec.\ \ref{sec:dres} we present the results of our DMRG calculations for the phase diagram. 
The results of the RKKY perturbation theory and the distance dependence of the RKKY coupling are presented in Sec.\ \ref{sec:rkky}. 
In Sec.\ \ref{sec:box} we discuss the relation to the quantum-box physics and the impact of a self-consistent treatment of the pairing strength for a simple toy model.
The main conclusions are summarized in Sec.\ \ref{sec:con}.

\section{Model and symmetries}  
\label{sec:mod}

We study systems with $M=2$ quantum-spin-$\frac12$ impurities $\ff S_{m}$ coupled via an antiferromagnetic local exchange of strength $J$ to a conventional s-wave BCS superconductor with gap $2\Delta$. 
The total Hamiltonian 
\be
H = H_{\rm hop} + H_{\rm ex} + H_{\rm sc}
\label{eq:ham}
\ee
includes the hopping term
\be
H_{\rm hop} = - t \sum_{i=1}^{L-1} \sum_{\sigma=\uparrow,\downarrow} 
\left(c_{i\sigma}^\dagger c_{i+1,\sigma} + c_{i+1,\sigma}^\dagger c_{i\sigma}\right)
\label{eq:hhop}
\: , 
\ee
where the nearest-neighbor hopping $t \equiv 1$ sets the energy scale, and where $c_{i\sigma}$ annihilates an electron at site $i=1,...,L$ with spin projection
$\sigma = \uparrow, \downarrow$.
We consider a one-dimensional chain of $L$ sites with open boundaries. 
The model is studied at half-filling, i.e., for total particle number $\langle N \rangle = L$, where $N = \sum_{i\sigma} c^{\dagger}_{i\sigma} c_{i\sigma}$ is the total particle-number operator.

The interaction of the impurity spins $\ff S_{m}$ with the local spin moments $\ff s_{i_{m}}$ of the electron system at the sites $i_{m}$ for $m=1,...,M$ of the lattice is given by
\be
H_{\rm ex} = J \sum_{m=1}^M \ff{S}_m \ff{s}_{i_m}
\; 
\label{eq:hex}
\ee
with $J>0$.
Here, $\ff s_{i} = \frac12\sum_{\sigma\sigma'} c_{i\sigma}^{\dagger} \ff \tau_{\sigma \sigma'} c_{i\sigma'}$, where $\ff \tau$ is the vector of Pauli matrices. 
For $M=2$ we will study geometries where the impurity spins are coupled to sites $i_{1}$ and $i_{2} > i_{1}$ at a distance $d = i_{2} - i_{1}$ and placed symmetrically with respect to the chain center, i.e., $i_1 + i_2 = L+1$.
Finally,
\be
H_{\rm sc} = \sum_{i=1}^L \left(\Delta \, c_{i\uparrow}^\dagger c_{i\downarrow}^\dagger + \Delta^{\ast} c_{i\downarrow} c_{i\uparrow}\right)
\label{eq:hsc}
\ee
describes a system in a conventional s-wave superconducting state.
We assume that $\Delta$ is real and positive, $\Delta > 0$. 
Note that this can always be achieved by the U(1) gauge transformation defined via $c_{i\sigma} \rightarrow \sqrt{\frac{\Delta^\ast}{\vert\Delta\vert}} \, c_{i\sigma}$.
We also have the usual assumption that $\Delta$ is spatially constant. 
This assumption can be relaxed slightly, as will be discussed later, along with an attempt for a self-consistent evaluation of the pairing strength $\Delta \mapsto \Delta_{i} = U \langle c_{i\downarrow} c_{i\uparrow} \rangle$ with an attractive Hubbard-$U$ term. 

The Hamiltonian is invariant under SU(2) spin rotations and thus conserves the total spin $\ff S \equiv \sum_{i} \ff s_{i} + \sum_{m} \ff S_{m}$. 
At $\Delta = 0$ and half-filling, there is an additional SU(2) symmetry in the charge sector (``hidden'' SU$_{\rm c}$(2) charge symmetry \cite{JV87}) generated by the components of the total pseudo-spin $\ff \eta \equiv \sum_{i} \ff \eta_{i}$, i.e., $[\ff \eta , H]=0$.
The $z$ component of the local pseudo-spin at site $i$ is defined as $\eta_{iz} = (c_{i\uparrow}^\dagger c_{i\uparrow} + c_{i\downarrow}^\dagger c_{i\downarrow} - 1)/2$, and the $x$ and $y$ components, $\eta_{ix} = \frac12 (\eta_{i+} + \eta_{i-})$ and $\eta_{iy} = \frac{1}{2i} (\eta_{i+} - \eta_{i-})$, are given in terms of $\eta_{i+} = \epsilon_i c_{i\uparrow}^\dagger c_{i\downarrow}^\dagger$ and $\eta_{i-} = \epsilon_i c_{i\downarrow} c_{i\uparrow}$. 
Here, $\epsilon_i = (-1)^{i}$ is the usual sign factor for a bipartite lattice.
The components of $\ff \eta_{i}$ span the charge $\mathfrak{su}(2)$ algebra with $[ \eta_{i\alpha} , \eta_{i\beta} ] = \sum_{\gamma=x,y,z} \epsilon_{\alpha\beta\gamma} \eta_{i\gamma}$.
We have $[s_{i\alpha} , \eta_{i'\alpha'}] = 0$ for arbitrary $\alpha,\alpha' = x,y,z$.  

Furthermore, if $L$ is even, an integer (half-integer) total spin $S$ implies an integer (half-integer) total pseudo-spin $\eta$, while for odd $L$
an integer (half-integer) total spin $S$ implies a half-integer (integer) total pseudo-spin $\eta$. 
Under antiunitary time reversal, $\ff s_{i} \mapsto - \ff s_{i}$ while $\ff \eta_{i} \mapsto (\eta_{ix}, -\eta_{iy}, \eta_{iz})$, and under unitary particle-hole transformation $\ca C=\prod (c_{i\sigma} - c_{i\sigma}^{\dagger})$, we have $\ff s_{i}  \mapsto (-s_{ix}, s_{iy}, - s_{iz})$ and likewise $\ff \eta_{i}  \mapsto (-\eta_{ix}, \eta_{iy}, - \eta_{iz})$. 
Finally, we have $\ff s_{i} \mapsto \ff \eta_{i}$ under $\ca C_{\downarrow} Z_{\downarrow}$, i.e., under the particle-hole transformation $\ca C_{\downarrow}$ in the spin $\downarrow$ sector combined with the staggered-sign transformation $Z_{\downarrow}$ for $\downarrow$ electrons defined via $c_{i\downarrow} \mapsto Z_{\downarrow} c_{i\downarrow} Z_{\downarrow}^{\dagger} = \epsilon_i c_{i\downarrow}$.

At finite $\Delta > 0$, and since 
\be
H_{\rm sc} = 2\Delta \sum_{i=1}^L \epsilon_i \eta_{ix}
\label{eq:hscd}
\ee
is proportional to the $x$ component of the total {\em staggered} pseudo-spin, the charge (pseudo-spin) SU(2) symmetry is broken.
This implies that $\eta_{z} = (N - L)/2$ and hence $N$ are no longer conserved. 
Only the pseudo-spin $U(1)$ symmetry around the $x$ axis and the conservation of $\eta_{x}$ (``pseudo-charge'') with (half-)integer eigenvalues $q$
remains \cite{YO00}. 
Furthermore, for $\Delta>0$ we still have conservation of the fermion parity $\Pi \equiv (-1)^{N}$. 
{\em Spatial} parity, unitarily represented by $P = \sum_{i\sigma} c_{i\sigma}^\dagger c_{L+1-i,\sigma}$ is conserved. 
For odd $L$ (only), $P$ additionally commutes with $\eta_{x}$.

\section{DMRG computations}  
\label{sec:dmrg}

To compute ground-state expectation values and correlations, we use a variational approach to optimize matrix-product states. 
Our algorithm is based on the ideas of DMRG \cite{Whi92,Sch11}. 
It relies on the finding that the ground state of gapped one-dimensional Hamiltonians with short-range hopping and interactions obeys an entropic area law, i.e., the entanglement entropy does not scale with the system size. 
Matrix-product states provide a well-suited platform for the handling of such low-entanglement quantum states.
Furthermore, our code fully exploits the non-abelian spin-SU(2) symmetry by using symmetry-reduced basis states \cite{McCG02,Wei12},  
and by using a lossless compression of matrix-product operators \cite{HMcCS17}. 
The bond dimension of symmetry-reduced state matrices must take values of the order of $10^3$ to provide a numerically precise approximation of the quantum states under study. 
With increasing $\Delta$, the required bond dimension becomes smaller, because the ground-state entanglement decreases when the correlations become predominantly local.
The convergence of the algorithm is controlled by the measure $\Delta E^2/L = (\langle H^2 \rangle - \langle H\rangle^2)/L$ and is additionally monitored by the change of the eigenvalues of the state matrices during the local optimization.
For all results presented, we have achieved a total-energy variance per site smaller than $\Delta E^2/L = 10^{-9}$.
The implementation of the spin-SU(2) symmetry allows us to target spin quantum numbers without numerical interference of energetically close states from other subspaces. 
This is important in order to distinguish between the ground-state energies for different spin quantum numbers with a high degree of accuracy.

\section{DMRG Results}
\label{sec:dres}

\subsection{Phase diagram for distance $d=1$}  

\begin{figure}[t]
\includegraphics[width=0.95\columnwidth]{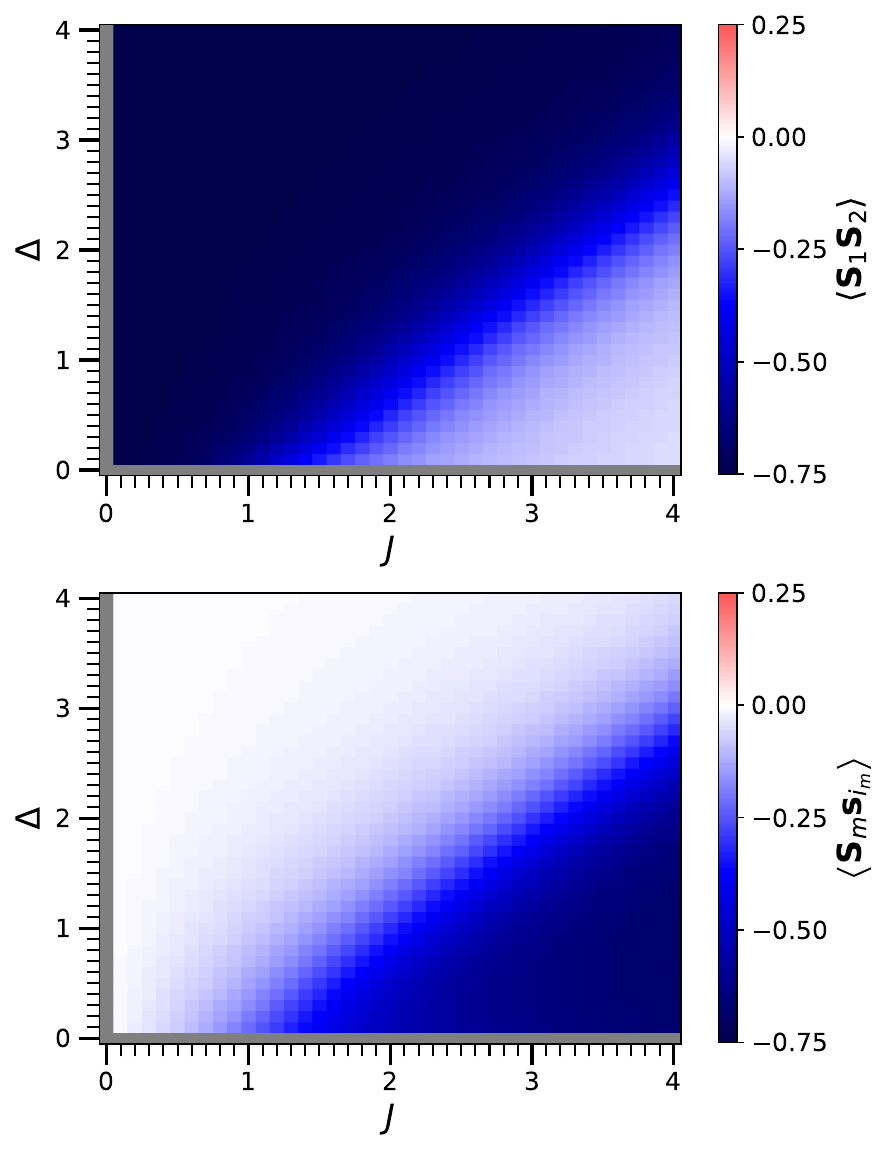}
\caption{
Impurity spin correlation $\left\langle\boldsymbol{S}_1\boldsymbol{S}_2\right\rangle$ (see color code, upper plot) and Kondo correlation function $\langle \ff S_{m} \ff s_{m} \rangle$ (lower plot)
as obtained by DMRG for a system consisting of two $s=\frac12$ impurity spins ($M=2$), exchange coupled to the nearest-neighbor ($d=1$) central sites of a one-dimensional chain with $L=82$ sites as function of $J$ and $\Delta$. 
The nearest-neighbor hopping $t=1$ fixes the energy scale.
}
\label{fig:nn}
\end{figure}

The competition between BCS-singlet formation, Kondo screening, and the emergent indirect RKKY exchange crucially depends on the distance $d$ between the impurity spins. 
For nearest neighbors, $d=1$, the upper plot in Fig.\ \ref{fig:nn} shows the impurity-spin correlation $\langle \ff S_{1} \ff S_{2} \rangle$ in the ground state of a long chain ($L=82$ sites). 
The impurity spins are coupled to sites in the center of the chain.

We find that the correlation is antiferromagnetic at weak $J$ and small $\Delta$, i.e., $\langle \ff S_{1} \ff S_{2} \rangle \approx - \frac34$. 
The correlation stays antiferromagnetic for any value of the pairing strength $\Delta$.
This is understood by RKKY-type perturbation theory, as discussed later in Sec.\ \ref{sec:rkky}.
Contrary, when increasing $J$, the RKKY singlet gradually breaks up, i.e., the absolute value of the correlation decreases and, in the ultimate strong-$J$ limit, $\langle \ff S_{1} \ff S_{2} \rangle \to 0$ (see the color code).

This transition from the weak to the strong-$J$ regime is a smooth crossover. 
In fact, the impurity-spin correlation varies smoothly as function of $J$ and $\Delta$ in the entire parameter space. 
This is consistent with the numerical finding that the ground state $| \Psi_{0} ( J, \Delta) \rangle$ is a total-spin singlet in the entire parameter space $(J, \Delta)$.
Note that this has been checked numerically by targeting ground states in sectors with total-spin quantum number $S > 0$ (in practice up to $S = \frac32$) in the calculations and comparing respective ground-state energies.
We have also traced the fermion parity, which is well defined in the non-degenerate ground state. 
$\Pi$ is even in the entire phase diagram.

However, the character of the singlet ground state is very different in the different limits. 
For $J \gg \Delta, t$ the ground state exhibits strong antiferromagnetic Kondo correlations $\langle \ff S_{m} \ff s_{i_{m}} \rangle \approx - \frac34$ ($m=1,2$) at the impurity sites, see lower plot in Fig.\ \ref{fig:nn}.
For $J \to \infty$, these essentially decouple the left and the right remainder of chain, $i < i_{1}$ and $i > i_{2}$, which then form two individual conventional BCS singlets.
Generally, the local Kondo correlations are strongest, $\langle \ff S_{m} \ff s_{i_{m}} \rangle \to - \frac34$, for strong $J$ and small $\Delta$.
Contrary, for weak $J$ and large $\Delta$, we have $\langle \ff S_{m} \ff s_{i_{m}} \rangle \to 0$. 
This is just complementary to the impurity-spin correlation $\langle \ff S_{1} \ff S_{2} \rangle$. 
The transition between these limits is featureless and smooth.

Our DMRG results are almost perfectly independent of the system size as long as $\Delta$ is larger than the finite-size gap $\delta \sim 1/L$. 
We found that this limits our studies of ground-state correlations to $\Delta \gtrsim 0.1$. 
Given this limitation, our results for $L=82$ are fully converged.
This is also true for all DMRG results throughout the paper.
We also note that the results for the correlation functions and, at finite $\Delta$, the total spin $S$ are independent of whether $L$ is selected as even or odd (while keeping $\langle N \rangle=L$).
The same holds for the case $d=2$ discussed below.

\subsection{Phase diagram for distance $d=2$}  

A nontrivial phase diagram is obtained for impurity spins coupled to next-nearest-neighbor sites ($d=2$). 
Figure \ref{fig:nnns} shows the $J$-$\Delta$ phase diagram for two impurity spins exchange coupled to the central sites $i_{1}=41$ and $i_{2}=43$ of a chain with $L=83$ sites.
Calculations for different $L$ and also for different positions $i_{1}$, $i_{2}$, but keeping $d=2$ constant, do not lead to any significant changes as long as the impurities are sufficiently far from the chain edges. 
Four different phases are found and discussed in the following, see Fig.\ \ref{fig:pd}.

\subsubsection{RKKY-triplet phase}

For weak $J$ and small $\Delta$, the ground state is a degenerate total-spin triplet $S=1$ (see Fig.\ \ref{fig:nnns} green color). 
Here, the two impurity spins are weakly coupled to the chain and form a nonlocal spin triplet while the host electron system is in a BCS singlet state. 
This spin triplet remains unscreened for the gapped system at $\Delta>0$, opposed to the $\Delta=0$ case, where screening takes place via two channels of the electron system \cite{NB80,ALJ95}.

The formation of an impurity-spin triplet for small $\Delta>0$ is the obvious explanation for the total-spin $S=1$ ground state: 
It relies on the well-known RKKY perturbation theory for the gapless ($\Delta=0$) case, which for a bipartite chain with nearest-neighbor hopping yields an effective RKKY interaction strength,
\be
J_{\rm RKKY} = - J^{2} \chi_{i_{1}i_{2}}(\omega=0)
\: , 
\label{eq:rkky}
\ee
which is negative, i.e., ferromagnetic ($H_{\rm eff} = J_{\rm RKKY} \ff S_{1} \ff S_{2}$), since the magnetic response given by the retarded static (frequency $\omega=0$) magnetic susceptibility is positive $\chi_{i_{1}i_{2}}(\omega=0)>0$ at distance $d=2$.
It will be shown, see our discussion below in Sec.\ \ref{sec:rkky}, that this also holds for small but finite $\Delta$.

The triplet formation is seen in our DMRG calculations of the impurity-spin correlation $\langle \ff S_{1} \ff S_{2} \rangle$ shown in the upper plot of Fig.\ \ref{fig:nnnss} for the same system. 
In fact, for $J \to 0$ and finite but sufficiently small $\Delta$, we get $\langle \ff S_{1} \ff S_{2} \rangle \to \frac14$, i.e., ferromagnetic correlation.
Furthermore, the lower plot in the figure demonstrates that the local Kondo correlations $\langle \ff S_{m} \ff s_{i_{m}} \rangle$ vanish in this regime.

\subsubsection{RKKY-singlet phase}

\begin{figure}[t]
\includegraphics[width=0.85\columnwidth]{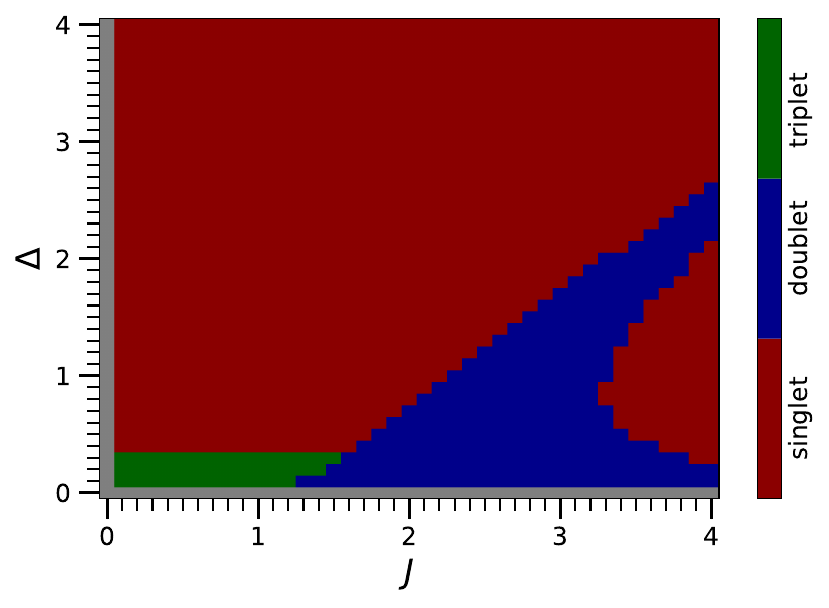}
\caption{
Total-spin quantum number $S$ for a system with $L=83$ lattice sites and two $s=\frac12$ impurity spins coupled to next-nearest-neighbor ($d=2$) central sites: $L=83$, $i_{1}= 41$, $i_{2}=43$.
Results as a function of $J$ and $\Delta$.
{\em Red:} $S=0$, {\em blue:} $S=\frac12$, {\em green:} $S=1$.
}
\label{fig:nnns}
\end{figure}

\begin{figure}[t]
\includegraphics[width=0.95\columnwidth]{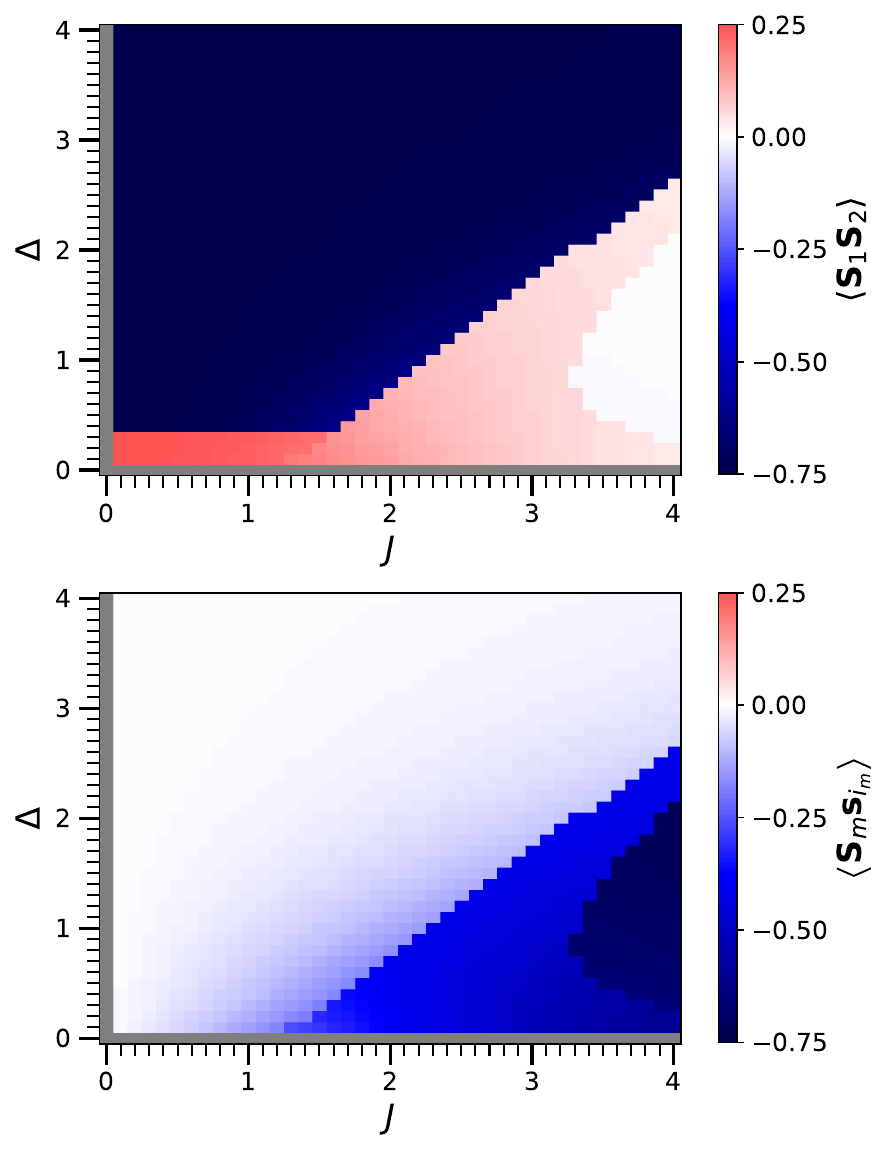}
\caption{
Impurity-spin correlation function $\langle \ff S_{1} \ff S_{2} \rangle$ (upper plot) and Kondo correlation function $\langle \ff S_{m} \ff s_{m} \rangle$ (lower plot) for $d=2$ (the same setup as for Fig.\ \ref{fig:nnns}). 
}
\label{fig:nnnss}
\end{figure}

Fig.\ \ref{fig:nnnss} also shows that for $\Delta > \Delta_{\rm c} \approx 0.35$ and $J\to 0$ the impurity-spin correlation changes its sign and immediately becomes strongly antiferromagnetic, i.e., $\langle \ff S_{1} \ff S_{2} \rangle \approx - \frac34$. 
Above the critical pairing strength $\Delta_{\rm c}$, and for weak $J$, the impurity spins form a spin-singlet state, which is disentangled from the BCS singlet of the electron system. 
Hence, the full system is in a total-spin singlet $S=0$ state. 
This RKKY-singlet phase is the second phase (ii) found in the diagram given by Fig.\ \ref{fig:nnns}. 

\begin{figure}[b]
\includegraphics[width=0.9\columnwidth]{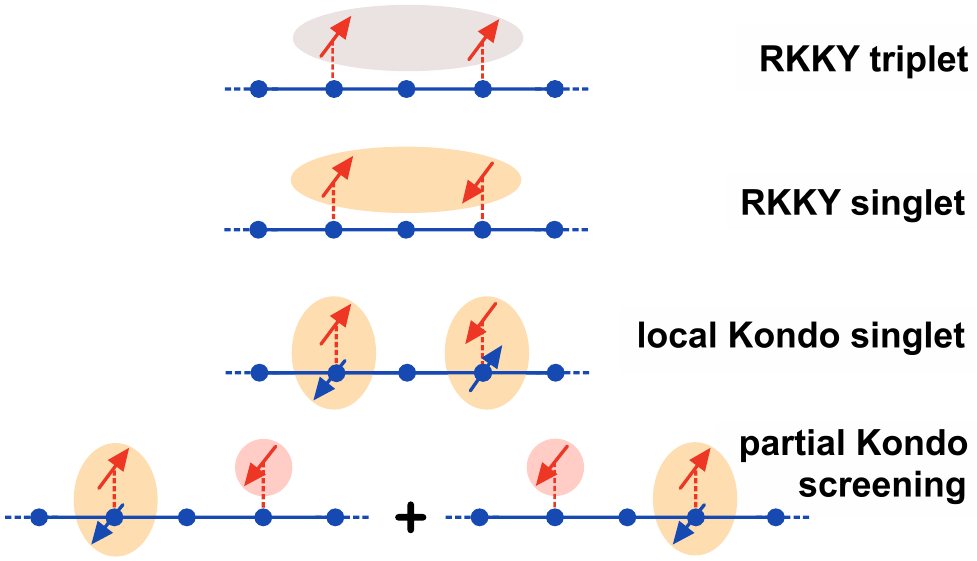}
\caption{
Sketch of the four different phases for distance $d=2$ between the $s=\frac12$ impurity spins.
Electron system: blue. Impurity spins: red. Singlets: light yellow. Triplet: light gray. Doublet: light red.
See text for discussion. 
}
\label{fig:pd}
\end{figure}

Both, the ground states of the RKKY triplet and the RKKY-singlet phase are sketched in Fig.\ \ref{fig:pd}. 
The transition can be understood as a discontinuous level crossing of subgap bound states in the spirit of Refs.\ \cite{ZLL+10,ZBP11,YMW+14}. 
We will demonstrate below (Sec.\ \ref{sec:rkky}) that, at weak $J$, the presence of the RKKY-singlet phase as well as the transition from the triplet to the singlet phase at $\Delta_{\rm c}$ can also be understood perturbatively. 

The RKKY-singlet phase extends to arbitrarily large $\Delta$ at weak $J$, and with increasing $J$ is bounded from below by larger and larger pairing strengths $\Delta$.
At the corresponding phase boundary $\Delta_{\rm c}(J)$, the impurity-spin correlation $\langle \ff S_{1} \ff S_{2} \rangle$ jumps from strongly antiferromagnetic (Fig.\ \ref{fig:nnnss}, upper plot, blue) to slightly ferromagnetic (light red) and, for $J\to \infty$, to almost vanishing values. 
For strong $J$, the transition line $\Delta_{\rm c}(J)$ in Fig.\ \ref{fig:nnnss}, actually reflects the (local) competition between Kondo-singlet and BCS-singlet formation, individually for each impurity.
In fact, $\Delta_{\rm c}(J)$ nestles up against the transition line $\Delta^{(1)}_{\rm c}(J)$ of the {\em single}-impurity ($M=1$) model with increasing $J$.
In the single-impurity model this line separates between the ``Kondo''-screened ($S=0$) and a state with unscreened ($S=\frac12$) impurity spin, and between odd and even fermion parity $\Pi$.
The single-impurity phase diagram is shown in the Appendix \ref{sec:kim}, see Fig.\ \ref{fig:kim}.

\subsubsection{Local Kondo-singlet phase}

For fixed $\Delta >0$ and in the limit $J\to \infty$, one expects that both impurity spins are individually screened, i.e., that two completely local Kondo singlets (LKS) are formed at $i_{1}$ and $i_{2}$ with strongly antiferromagnetic and local Kondo correlations $\langle \ff S_{m} \ff s_{i_{m}} \rangle \to -\frac34$ (for $m=1,2$).
In fact, this can be seen in the lower plot of Fig.\ \ref{fig:nnnss} (dark blue).
Consequently, the impurity-spin correlation vanishes $\langle \ff S_{1} \ff S_{2}\rangle \to 0$ as $J \to \infty$ and for arbitrary values of $\Delta>0$ (upper plot of Fig.\ \ref{fig:nnnss}, white).

The two LKSs cut the state of the one-dimensional electron system into three fully disentangled spin-singlet states: 
The singlet at the site $i_{1}+1$, i.e., between the LKSs at $i_{1}$ and at $i_{2} = i_{1}+2$, is given by the eigenstate of the $x$ component of the local pseudo-spin $\eta_{ix}$, i.e., by the pseudo-spin-$\eta_{ix}=\frac12$ state $(| {\rm vac.} \rangle - |\uparrow \downarrow \rangle) / \sqrt2$. 
This state and the state $(| {\rm vac.} \rangle + |\uparrow \downarrow \rangle) / \sqrt2$ belong to a pseudo-spin doublet, which is split due to the finite ``field'' $\Delta$ locally coupling to $\eta_{ix}$.
The pseudo-spins of the two other spin singlets supported by $i=1,...,i_{1}-1$ and by $i=i_{2}+1,...,L$ depend on the lengths of the respective support. 
Here, for $i_{1}=41$ and $i_{2}=43$, both consist of an even number of sites. 
This implies an integer and, in fact, zero pseudo-spin. 
For an odd number of sites, one would have $\eta_{\rm left} = \eta_{\rm right}=\frac12$.

As can be seen in the upper and lower plot of Fig.\ \ref{fig:nnnss}, the singlet LKS phase (white and dark blue, respectively) is found at finite $J$ for pairing strengths $\Delta$ bounded from above by a second transition line $\Delta'_{\rm c}(J)$ (with $\Delta'_{\rm c}(J) < \Delta_{\rm c}(J)$). 
It is bounded from below by a third line $\Delta_{\rm c}''(J)$. 
With increasing $J$ the range of pairing strengths, $\Delta'_{\rm c}(J) < \Delta < \Delta''_{\rm c}(J)$, in which the LKS phase is realized, grows, i.e., 
$\Delta'_{\rm c}(J)$ increases and $\Delta''_{\rm c}(J)$ decreases with $J$, eventually extending over the entire $\Delta$ axis in the phase diagram.

\subsubsection{Partial Kondo-screened phase}

The remaining doublet phase in the phase diagram in Fig.\ \ref{fig:nnns} (blue color) results as a compromise between the $J$ driven tendency to form two LKSs, realized for $\Delta < \Delta'_{\rm c}(J)$, and the $\Delta$ driven formation of a more delocalized BCS singlet state with a nonlocal RKKY two-impurity spin singlet, realized for $\Delta > \Delta_{\rm c}(J)$. 
The competition and the resulting compromise necessarily requires two impurity spins and is meaningless in the single impurity case. 
We refer to the state as partial Kondo screening (PKS).

In fact, in the ground state for $\Delta'_{\rm c}(J) < \Delta < \Delta_{\rm c}(J)$, only a single impurity spin, say $\ff S_{1}$, undergoes a local Kondo screening, while the second ($\ff S_{2}$) remains unscreened and is responsible for the twofold ground-state degeneracy with total-spin quantum number $S=\frac12$.
At the site $i_{1}$ a single electron is localized, and the local electron spin moment $\langle \ff s_{i_{1}}^{2} \rangle \to \frac34$ deep in the PKS limit, i.e., for $J\to \infty$ and $\Delta'_{\rm c}(J) < \Delta < \Delta_{\rm c}(J)$, while the local moment at $i_{2}$ is suppressed $\langle \ff s_{i_{2}}^{2} \rangle \to 0$.
The true ground state in the PKS limit is a superposition with a second state, where the roles of $\ff S_{1}$ and $\ff S_{2}$ are interchanged, i.e., where $\ff S_{2}$ is screened while $\ff S_{1}$ is decoupled from the rest of the system. 
The bonding ground state $\ket \Psi_{+}$ and the antibonding excited state $\ket \Psi_{-}$ thus approach the form
\ba
\ket \Psi_{\pm} 
&= 
\frac{1}{\sqrt2} \Big(
\ket{\mathrm{BCS}}_{\overline{1}} \otimes \ket{\mathrm{LKS}}_{1} \otimes \ket{M}_{2}
\nonumber \\
&\pm 
\ket{\mathrm{BCS}}_{\overline{2}} \otimes \ket{M}_{1} \otimes \ket{\mathrm{LKS}}_{2}
\Big)
\: , 
\label{eq:pks}
\ea 
where the BCS singlet states, labelled by $\overline{m}$, extend over all sites except for the site $i_{m}$, and where $| M=\pm \frac12 \rangle_{m}$ are the eigenstates of $S_{mz}$.
The degenerate ground-state spin doublet $\ket \Psi_{+}$ has odd spatial parity $p=-$, and the states $\ket \Psi_{-}$ forming the antibonding doublet have even parity $p=+$.
Eq.\ (\ref{eq:pks}) immediately implies that $\langle \ff S_{1} \ff S_{2} \rangle = 0$, consistent with the DMRG result in the deep PKS limit, see Fig.\ \ref{fig:nnnss} (upper plot). 
Also the local Kondo correlation in the ground state is easily computed from Eq.\ (\ref{eq:pks}) and turns out as $\langle \ff S_{m} \ff s_{m} \rangle = -\frac38$ for $m=1,2$. 
Again, this is in perfect agreement with the numerical result, see the lower plot in Fig.\ \ref{fig:nnnss}. 

For weak $J$ but still in the PKS phase with $\Delta < \Delta_{\rm c}(J)$, ferromagnetic impurity-spin correlations build up, as is seen in Fig.\ \ref{fig:nnnss} (upper plot) around $\Delta \approx 0.2$.
Across the transition from the PKS to the ferromagnetic triplet phase for weak $J$, $\langle \ff S_{1} \ff S_{2} \rangle$ evolves smoothly. 
Note that we continue to refer to the phase under consideration as ``PKS'' for simplicity, even though Eq.\ (\ref{eq:pks}) no longer holds.

For small $\Delta>0$ and strong $J$, but still within the PKS phase, the ground state no longer has the form (\ref{eq:pks}): 
In the limit $J\to \infty$, our numerical results still yield $\langle \ff S_{1} \ff S_{2} \rangle = 0$, but $\langle \ff S_{m} \ff s_{m} \rangle \to - \frac34$, i.e., we find the same correlations as in the LKS state. 
For strong $J$ and across the transition from the PKS to the LKS phase, $\langle \ff S_{m} \ff s_{m} \rangle$ is developing more and more smoothly. 

For small $\Delta>0$ and weak $J$ within the PKS phase, on the other hand, the situation is less clear. 
The data could be consistent with a critical pairing strength $\Delta_{\rm c}(J) \to 0$ for $J\to 0$ (see Fig.\ \ref{fig:nnns}), such that the PKS state with $\Delta<\Delta_{\rm c}(J)$ terminates exactly at the point $\Delta=0$ and $J=0$ in the phase diagram with correlations $\langle \ff S_{1} \ff S_{2} \rangle \to \frac14$ and $\langle \ff S_{m} \ff s_{m} \rangle \to 0$ (cf.\ Fig.\ \ref{fig:nnnss}), thereby approaching those of the RKKY-triplet phase. 
A direct numerical proof, however, would require calculations for $\Delta \ll 0.1$ and correspondingly for much larger values of $L$.
We will revisit the discussion of this parameter regime in Sec.\ \ref{sec:box}.

\subsection{Discussion}

Partial Kondo screening, i.e., screening of only a single spin in a unit cell, has been discussed in the context of the magnetically frustrated Kondo-{\em lattice} \cite{MNYU10}, for the Anderson model \cite{HUM11,AAP15} on the triangular lattice, and the Kondo lattice on the zigzag ladder \cite{PRP18,PWP19}.
As it emerges as a result of the competition between RKKY coupling and Kondo screening, the PKS phase is typically located between an (RKKY driven) magnetic phase and a paramagnetic heavy-fermion phase with strong local Kondo correlations in the respective ground-state phase diagrams.

In the context of the two-impurity-spin model, the PKS ground state was first suggested in the NRG study, Ref.\ \cite{YMW+14} (and called ``molecular doublet''). 
However, there is a qualitative difference between the more generic NRG phase diagram for a flat density of states and treating $J_{\rm RKKY}$ as an independent parameter rather than an emergent coupling, opposed to the present phase diagram for the concrete one-dimensional geometry: 
In the phase diagram of Ref.\ \cite{YMW+14}, the PKS phase is ``detached'' from the $\Delta=0$ line and extends over a limited parameter space only. 
Contrary, in Fig.\ \ref{fig:nnns}, the phase appears to merge with the gapless $\Delta = 0$ state in a rather wide range of exchange interactions $J$. 

In the gapless model at $\Delta=0$, we do not expect a quantum-critical point at a finite $J$ but rather a smooth crossover from the RKKY to the Kondo regime with increasing $J$, see Refs.\ \cite{JVW88,JV89,SS92,Fye94,ALJ95,SLO+96} and the discussion in Ref.\ \cite{ELA18}.

Another interesting question is, what the phase diagrams for larger $d$ look like and how the $d\to \infty$ limit is approached.
Obviously, as the distance $d$ between the impurity spins increases, the physical properties are more and more governed by the physics of the single-impurity model (see Appendix \ref{sec:kim}). 
We have computed the $J$-$\Delta$ phase diagrams for larger distances $d =3$ and $d=4$, to explore how the single-impurity physics is approached. 
It turns out that the trend is non-monotonic but strictly alternating between odd and even $d$. 
This reflects the bipartite geometry of the model system studied.

In fact, the topology of the phase diagrams for odd $d$ and for even $d$ remains the same as $d$ increases.
In particular, there are smooth crossovers rather than first-order transitions between the different regimes for odd $d$, while for even $d$ the phase diagrams consist of the above-discussed four different phases. 
The most important quantitative effect is the exponentially strong decrease of $J_{\rm RKKY} = J_{\rm RKKY}(d)$ with $d=i_{1} - i_{2}$ for large $d$.
Hence, the single-impurity limit is approached rapidly as $d$ increases. 
This also crucially depends on $\Delta$, as the $\Delta$ dependence of $J_{\rm RKKY} = J_{\rm RKKY}(\Delta)$ is exponential as well, for large $\Delta$ (cf.\ Sec.\ \ref{sec:rkky}).
Furthermore, from the comparison between $d=2$ and $d=4$ (not shown), we conclude that the phase space taken by the PKS phase at even $d$ shrinks to zero as $d \to \infty$. 

Finally, it is noticeable that the $d=1$ phase diagram actually consists of a single phase only and that, opposed to $d=2$, there are smooth crossovers only rather than discontinuous transitions.
Key to the qualitatively rather different phase diagrams is the pseudo-charge $q$, i.e., the eigenvalue of $\eta_{x}$. 

Consider the deep PKS regime and the ground state $\ket \Psi_{+} = \ket \Psi_{1} + \ket \Psi_{2}$, see Eq.\ (\ref{eq:pks}), where $\ket \Psi_{m}$ for, say, $m=1$, refers to the state where $\ff S_{1}$ and $\ff s_{i_{1}}$ form a local Kondo singlet, while $\ff S_{2}$ remains unscreened.
In the state $\ket \Psi_{1}$ the pseudo-charge $q_{i_{1}}=0$, locally at site $i_{1}$, while the pseudo-charge at $i_{2}$ is finite, namely $q_{i_{2}}= \pm \frac12$, and induced by the conjugate ``field'' $\propto (-1)^{i_{2}} \Delta$, see Eq.\ (\ref{eq:hscd}).

If the distance is $d=2$, and since the field is alternating due to the sign factor $\epsilon_{i}=(-1)^{i}$ in Eq.\ (\ref{eq:hscd}), the sum of the local pseudo-charges at the three sites $i_{1}$, $i_{1}+1$, and $i_{2}$ is zero for $\ket \Psi_{1}$ and for $\ket \Psi_{2}$, and thus for $\ket \Psi$.  
Contrary, if $d=1$, the pseudo-charges at $i_{1}$ and $i_{2}$ add to $\pm \frac12$ for $\ket \Psi_{1}$ and to $\mp \frac12$ for $\ket \Psi_{2}$. 
This implies that the PKS state $\ket \Psi_{1} + \ket \Psi_{2}$ cannot be an eigenstate of $H$.
Qualitatively, this explains the absence of a PKS phase in the $d=1$ phase diagram.
Furthermore, for $d=1$ the RKKY coupling is antiferromagnetic. 
Hence, in the whole parameter space the ground state is expected to be a total-spin singlet state with a fixed pseudo-charge. 
The latter is $q=0$ or $q=\pm \frac12$ for odd or even $L$, respectively, and is the same in the RKKY and the LKS regimes.

For the gapped system considered here, discontinuous transitions result from level crossings between different subgap bound states upon variation of the model parameters.
However, a crossing between bound states in the same $(S,q)$ sector requires the fine-tuning of three independent parameters \cite{vNW29}. 
From this we conclude that level transitions in the $J$-$\Delta$ plane are not to be expected and that the phase diagram therefore only shows smooth crossovers (cf.\ Fig.\ \ref{fig:nn}). 

\section{Perturbation theory}
\label{sec:rkky}

At any fixed $\Delta >0$ and for weak local exchange $J$, degenerate perturbation theory in $J$ applies and can be used to understand the occurrence of the RKKY triplet and the RKKY-singlet phase in the phase diagram (Fig.\ \ref{fig:nnns}) for $d=2$ at weak $J$, and the transition at $\Delta_{\rm c}(J) \approx 0.35$ in particular. 

Our calculations are done for the Hamiltonian (\ref{eq:ham}), i.e., a chain of length $L$, nearest-neighbor hopping $t$, finite $\Delta>0$, and $M$ spins $s=\frac12$ and make use of the fact that the BCS-type model $H_{\rm BCS} = H_{\rm hop} + H_{\rm sc}$ can be diagonalized analytically, see Appendix \ref{sec:bcs}. 
The results of the subsequent RKKY-type second-order perturbation theory in $J$, see Appendix \ref{sec:sopt}, can be summarized with the effective impurity-spin Hamiltonian
\be
  H_{\rm eff} = P_0 \sum_{mm'} J_{i_{m}i_{m'}} \ff S_m \ff S_{m'} P_0
\: . 
\ee
Here, $P_{0}$ is the projector onto the $2^{M}$-dimensional subspace of ground states at $J=0$. 
For $M=2$ quantum spins, we have $H_{\rm eff} = P_0 J_{\rm RKKY} \ff S_1 \ff S_{2} P_0$ with $J_{\rm RKKY} = 2 J_{i_{1}i_{2}}$ (the factor 2 is present for $i_{1} \ne i_{2}$ only). 

\begin{widetext}
For arbitrary $M$, the effective exchange couplings $J_{i_{m}i_{m'}} = J_{i_{m'}i_{m}}$ are obtained as (see Appendix \ref{sec:sopt}):
\begin{align}
&J_{i_{m}i_{m'}}
=
\frac{-J^2}{2(L+1)^2}\sum_{n,n'}\sin(i_mk_n)\sin(i_{m'}k_n)\sin(i_mk_{n'})\sin(i_{m'}k_{n'}) \: \frac{\varepsilon_n\varepsilon_{n'} - 4t^2\cos(k_n)\cos(k_{n'}) - \Delta^2}{\varepsilon_n\varepsilon_{n'}(\varepsilon_n+\varepsilon_{n'})}
\: .
\label{eq:jrkky}
\end{align}
Here, $\pm\varepsilon_n=\pm\sqrt{4t^2\cos^2(k_n)+\Delta^2}$ for $k_{n}=\frac{\pi n}{L+1}$ and $n=1,...,L$ are the one-particle energies. 
Both, the positive and the negative eigenenergies are twofold degenerate each, except for odd $L$ at $n=(L+1)/2$, where 
$\pm\varepsilon_n=\pm\Delta$.
We note that the limit $\Delta \to 0$ in Eq.\ (\ref{eq:jrkky}) is regular and that the resulting coupling constant $\lim_{\Delta\to 0} J_{\rm RKKY}(\Delta)$ equals the standard expression for $J_{\rm RKKY}$ that is obtained by perturbation theory in $J$, but starting from the $J=\Delta=0$ ground state as a reference, see the supplemental material of Ref.\ \cite{SGP12}.
\end{widetext}

\begin{figure}[b]
\includegraphics[width=0.95\columnwidth]{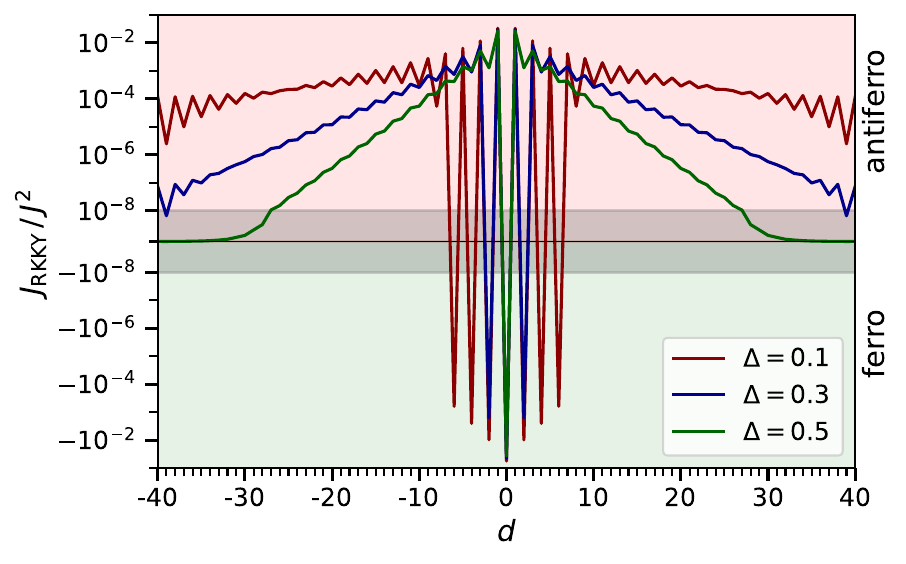}
\caption{
RKKY coupling $J_{\rm RKKY}$ as obtained from Eq.\ (\ref{eq:jrkky}) as a function of the distance $d$ between the impurity spins for various $\Delta$. 
Calculations for $L=81$ sites,
and impurities placed symmetrically with respect to the chain center at $i_{1}=41$ and $i_{2}=i_{1}+d$.
Note the log-linear scale: in the grey-shaded range, the scale is linear, else logarithmic.
The nearest-neighbor hopping $t\equiv 1$ sets the energy scale.
}
\label{fig:jofd}
\end{figure}

Fig.\ \ref{fig:jofd} shows the RKKY interaction strength as function of the distance $d$ between the impurity spins.
For $\Delta=0.1$ (red line), $J_{\rm RKKY}$ oscillates between ferromagnetic ($J_{\rm RKKY}<0$) and antiferromagnetic ($J_{\rm RKKY}>0$) when the distance is increased in units of one, and its magnitude decreases with $d$. 
This is very similar to what is known from RKKY theory for the gapless system at $\Delta=0$.
However, for $d \ge 7$, the distance dependence changes qualitatively: 
$J_{\rm RKKY}$ remains antiferromagnetic and decreases exponentially (note the log-linear scale in the figure).
In addition, there are some superimposed small oscillations with $d$, which fade out and essentially disappear for $d \approx 25$.
The $d$ dependence beyond $d \approx 25$ is no longer characteristic for the system in the thermodynamic limit, i.e., the growth of the oscillation amplitude with $d$ must be seen as an unwanted finite-size effect.

For a gapped system, an exponential decrease of $J_{\rm RKKY}(d) \propto \exp(- d / \xi)$ with a length scale $\xi \propto 1/\Delta$ for large $d$ and $\Delta$ is simply the result of the tunnelling effect \cite{BR55}. 
On the other hand, for small $d$ and $\Delta$, the contributions of high-energy excitations to $J_{\rm RKKY}$ dominate and cause oscillations as in the standard RKKY theory of gapless systems.
Furthermore, and plausibly, the absolute magnitude of RKKY coupling is at a maximum for $d=0$ and ferromagnetic (both spins couple to the same site) for any $\Delta$.
For $\Delta \to \infty$, we have the simple analytical result $J_{\rm RKKY}(d=0) = - J^{2} / 8 \Delta^{2}$.
Combined with the general sum rule $\sum_{i'=1}^{L} J_{ii'}=0$, valid for a singlet ground state, this implies that for large $d$ the coupling must be antiferromagnetic (positive coupling) to compensate for the strongly negative local term.
If $\Delta$ is increased, see blue line in Fig.\ \ref{fig:jofd}, the tunnel regime is reached earlier, i.e., at smaller $d$. 
For $\Delta = 0.5$ (green line), the RKKY coupling is antiferromagnetic for all distances $d$, except for $d=0$.
This also holds for larger $\Delta$.

\begin{figure}[t]
\includegraphics[width=0.95\columnwidth]{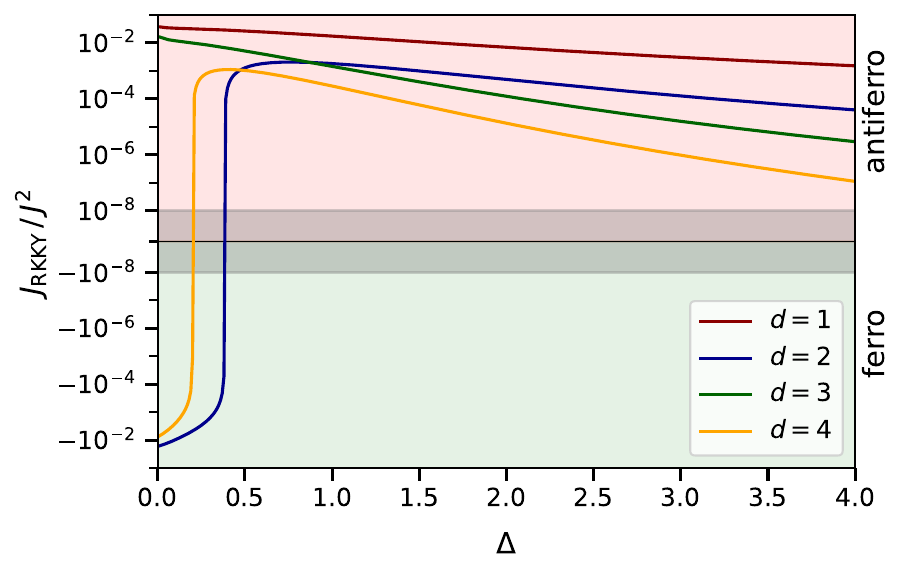}
\caption{
RKKY coupling $J_{\rm RKKY}$ as obtained from Eq.\ (\ref{eq:jrkky}) as a function of $\Delta$ for various $d$. 
Calculations for $L=81+d$ sites and impurities placed symmetrically with respect to the chain center at $i_{1}=41$ and $i_{2}=i_{1}+d$.
}
\label{fig:jofdel}
\end{figure}

For all even distances $d \ne 0$ this implies that the RKKY coupling must switch from ferromagnetic at small $\Delta$ to antiferromagnetic at large $\Delta$. 
This is demonstrated with Fig.\ \ref{fig:jofdel}, where the $\Delta$ dependence of $J_{\rm RKKY}$ is shown for $d=1,...,4$.
For $d=1$ the coupling is antiferromagnetic for any $\Delta$ and decreases exponentially.
The same behavior is found for $d=3$, except that $J_{\rm RKKY}$ is already considerably weaker. 
If $d$ is even, the RKKY interaction is ferromagnetic for small $\Delta$, and this explains the presence of the RKKY-triplet phase (see Fig.\ \ref{fig:nnns} for $d=2$).

However, as can be seen in the figure, there is a critical pairing strength $\Delta_{\rm c}(d)$, beyond which $J_{\rm RKKY}$ becomes positive such that the impurity spins couple antiferromagnetically.
For $d=2$ (see the blue line in Fig.\ \ref{fig:jofdel}), we have $\Delta_{\rm c} \approx 0.38$, consistent with the transition point ($\Delta_{\rm c} \approx 0.35$) seen in the phase diagram in Fig.\ \ref{fig:nnns} between the RKKY triplet and the RKKY singlet for $J\to 0$, given the chosen discretization of the $\Delta$ axis.
This critical pairing strength decreases with increasing $d$ (compare with the yellow line).
Fig.\ \ref{fig:jofdel} also demonstrates that $J_{\rm RKKY}$ decreases exponentially with increasing and sufficiently large $\Delta$ and for any given distance $d$. 
In the tunnel regime for large $d$ and $\Delta$ we have $J \propto e^{- \rm{const.} \Delta \, d}$.

To check the validity range of the perturbative approach, we have compared the normalized RKKY interaction $J_{\rm RKKY}/J^{2}$ with the DMRG results for different values of the local exchange coupling $J$.
The DMRG data are obtained from two independent computations targeting the $S=0$ and the $S=1$ sectors, respectively.
The RKKY coupling is then obtained as the total ground-state energy difference
\be
  J^{\rm (DMRG)}_{\rm RKKY} = E_0(S=1) - E_0(S=0)
  \: .
\ee
Perfect agreement between perturbation theory and the numerically exact DMRG can be expected only, if the ground states in the $S=0$ and $S=1$ sectors are given by a tensor product of a total impurity-spin singlet (triplet) state with the ground state of the purely electronic rest of the system: $\ket{\Psi_0(S)} = \ket{\rm imps.}_{S} \otimes \ket{\rm elect.}$. 

\begin{figure}[t]
\includegraphics[width=0.95\columnwidth]{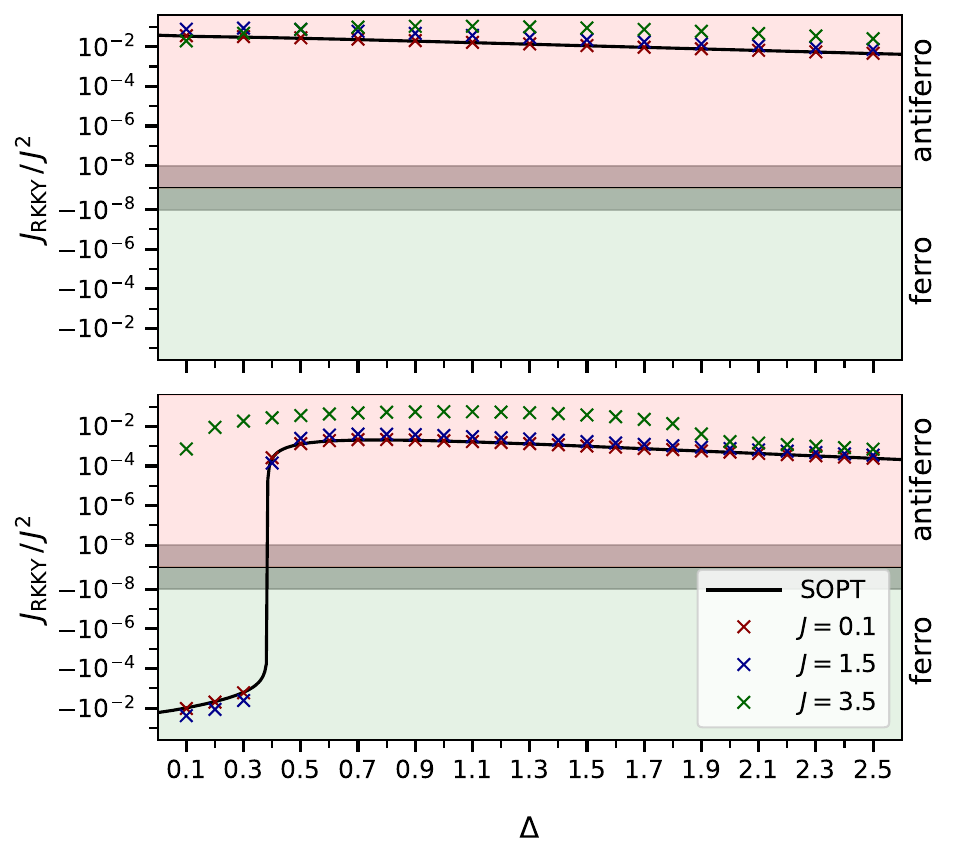}
\caption{
Normalized RKKY interaction $J_{\rm RKKY} / J^{2}$ as a function of $\Delta$. 
{\em Upper panel:} $d=1$ ($L=82$, $i_{1}=41$, $i_{2}=42$). 
{\em Lower panel:} $d=2$ ($L=83$, $i_{1}=41$, $i_{2}=43$).
Lines: RKKY second-order perturbation theory (SOPT).
Crosses: DMRG results for various values of the local exchange coupling $J$
}
\label{fig:comp}
\end{figure}

As is demonstrated with Fig.\ \ref{fig:comp}, the perturbative result for the $\Delta$ dependence of $J_{\rm RKKY}/J^{2}$, on a logarithmic scale, agrees quite well with the DMRG data.
Consider first the large-$\Delta$ regime. 
For both, $d=1$ (upper panel) and $d=2$ (lower panel), the DMRG results show a significant $J$ dependence, but $J^{\rm (DMRG)}_{\rm RKKY}/J^{2}$ monotonically decreases with decreasing $J$ and approaches the perturbative result.

For smaller $\Delta$ and at $J=3.5$ the perturbative result is no longer consistent with the DMRG data for both, $d=1$ and $d=2$. 
In particular, there is a rather strong deviation between the DMRG data and the perturbative result for $d=2$, at $J=3.5$ and for $\Delta \lesssim 1.9$ (see green crosses in the lower panel).
This is explained by the fact that local-Kondo-singlet formation becomes more and more dominant and that the impurity spins get more and more entangled with the electron system. 

Remarkably, for weak $J$ there is an almost perfect agreement between perturbation theory and DMRG for the critical pairing strength, at which the singlet and the triplet state become degenerate.
Consistent with the phase diagram in Fig.\ \ref{fig:nnns}, the transition at $\Delta_{\rm c} \approx 0.38$ from the RKKY singlet ($\Delta>\Delta_{\rm c}$) to the RKKY triplet ($\Delta<\Delta_{\rm c}$) is almost $J$ independent.
The nature of this phase transition is thus entirely accessible by perturbative means.

It is instructive to compare the results for quantum impurity spins $\frac12$ with approaches, where the impurity spins are treated as classical \cite{AMY97,GL02}, and where the crossover from oscillatory distance dependence of the inter-impurity coupling, at small $d$ and small $\Delta$, to antiferromagnetic coupling and exponential decay, at large $d$ and $\Delta$, is explained perturbatively in $J$. 
In Ref.\ \onlinecite{YGD+14} it is furthermore demonstrated by nonperturbative means that the presence of YSR bound states \cite{Yu65,Shi68,Rus69} and their $J$-dependent position in the gap pushes this crossover to significantly smaller distances.
Here, we find the same enhanced antiferromagnetic exchange within DMRG for the quantum-spin case. 
By comparison with the perturbative RKKY results and due to the almost perfect agreement with the DMRG data, we conclude that there is no particularly nonperturbative mechanism dominating over the perturbative RKKY coupling, at least for the parameter ranges considered.
However, consistent with Ref.\ \onlinecite{YGD+14}, there is a nonperturbative contribution to the enhancement of the antiferromagnetic exchange, i.e., $J^{\rm (DMRG)}_{\rm RKKY}/J^{2}$ is $J$ dependent and stronger for stronger $J$, see the results for $J=0.1$ and $J=1.5$ in Fig.\ \ref{fig:comp} for $\Delta > 0.5$.

\section{Quantum box}
\label{sec:box}

For small $\Delta \lesssim 0.1$, due to the growth of the entanglement entropy with decreasing gap size $2\Delta$, the DMRG calculations become increasingly costly.
We thus discuss a toy model with the same Hamiltonian, Eq.\ (\ref{eq:ham}), but on a small chain with $L=3$ sites only, for which the phase diagram can be computed by exact diagonalization. 
The two impurity spins are placed at distance $d=2$ and symmetrically around the central site, i.e., they are coupled to the sites $i_{1}=1$ and $i_{2}=3$. 
Opposed to the previously discussed systems, finite-size effects can no longer be neglected and actually dominate the physical properties, except for very large $\Delta$, where the superconducting gap exceeds the finite-size gap $\delta$ by far.
This ``quantum-box'' setup is interesting since, conceptually, it approaches the phase diagram of the system in the thermodynamical $L\to \infty$ limit from a different side and thus gives us a complementary perspective. 

\subsection{Exact-diagonalization results}  
\label{sec:ed}

\begin{figure}[t]
\includegraphics[width=0.85\columnwidth]{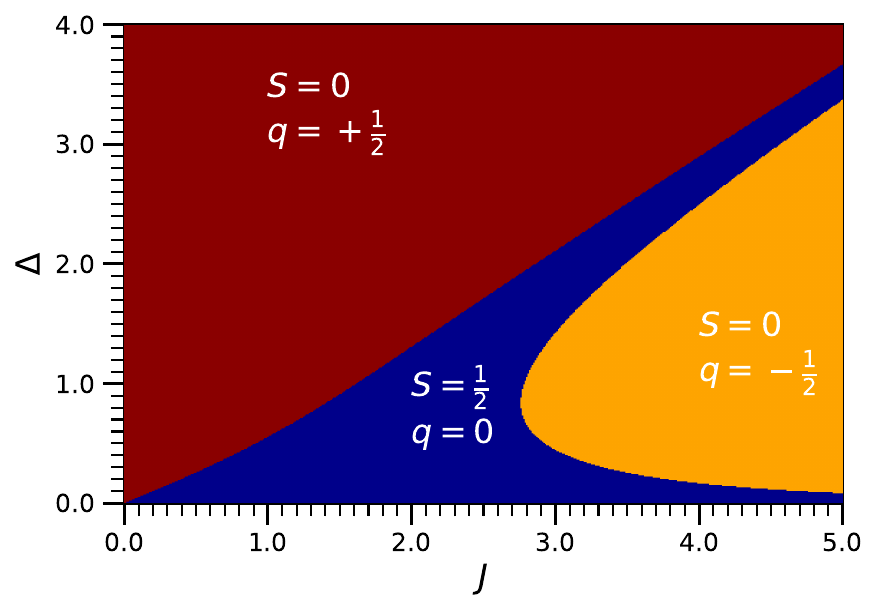}
\caption{
$J$-$\Delta$ ground-state phase diagram for distance $d=2$, as in Fig.\ \ref{fig:nnns}, but for $L=3$ sites ($i_{1}=1$, $i_{2}=3$).
Results for the total-spin quantum number $S$ and the pseudocharge $q$ as obtained by exact diagonalization.
}
\label{fig:toy}
\end{figure}

\begin{figure}[t]
\includegraphics[width=0.99\columnwidth]{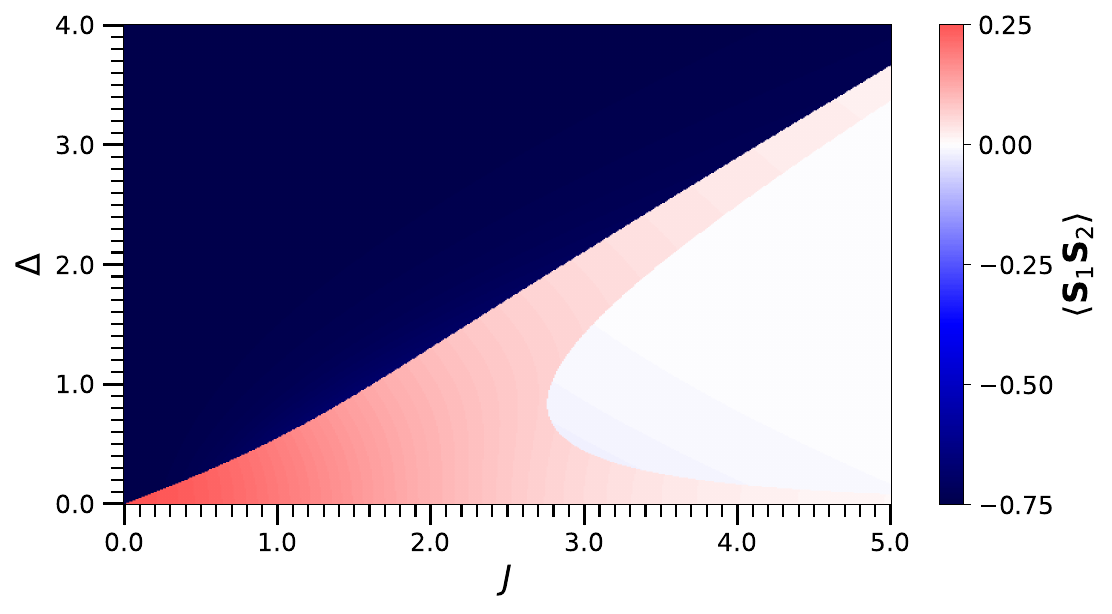}
\caption{
The same as Fig.\ \ref{fig:toy} but for the impurity-spin correlation function $\langle \ff S_{1} \ff S_{2} \rangle$ (color code).
}
\label{fig:toyss}
\end{figure}

\begin{figure}[t]
\includegraphics[width=0.99\columnwidth]{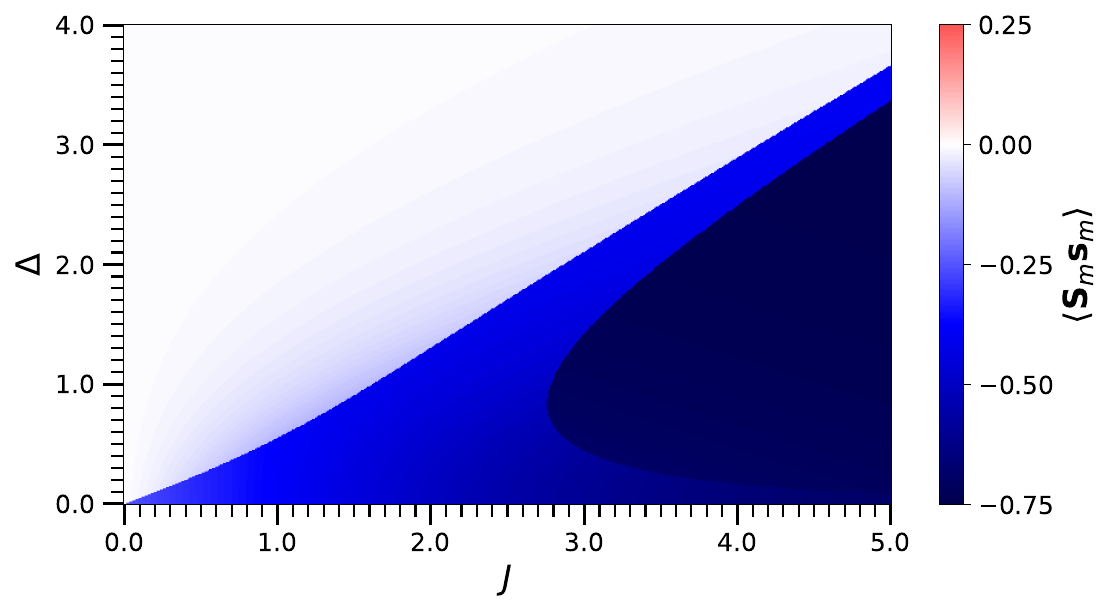}
\caption{
The same as Fig.\ \ref{fig:toy} but for the Kondo correlation function $\langle \ff S_{m} \ff s_{m} \rangle$ (color code).
}
\label{fig:toykondo}
\end{figure}

For the $L=3$ model, Fig.\ \ref{fig:toy} displays the boundaries between ground states with different total-spin quantum numbers. 
Figs.\ \ref{fig:toyss} and \ref{fig:toykondo} show the impurity-spin correlation $\langle \ff S_{1} \ff S_{2} \rangle$ and the local Kondo correlation $\langle \ff S_{m} \ff s_{i_{m}} \rangle$, respectively.
For strong $J$ and for $\Delta$ with $\Delta'_{\rm c}(J) < \Delta < \Delta_{\rm c}(J)$, the ground state is a degenerate spin doublet ($S=\frac12$), see blue region in Fig.\ \ref{fig:toy}.
For large $\Delta \to \infty$, it is of the PKS form, Eq.\ (\ref{eq:pks}), with correlations $\ff S_{1} \ff S_{2} \to 0$ and $\ff S_{m} \ff s_{i_{m}} \to - \frac38$ (see Figs.\ \ref{fig:toyss} and \ref{fig:toykondo}).
This deep PKS regime in the parameter space perfectly matches with the corresponding one for large $L$ ($L=83$, see Fig.\ \ref{fig:nnns}). 
This is due to the fact that correlations are almost completely local for strong $J$ and for $\Delta \gg \delta$. 

The orange-colored phase in Fig.\ \ref{fig:toy} is an LKS phase with $\ff S_{m} \ff s_{i_{m}} \to - \frac34$ for $J\to \infty$, and the red-colored one an RKKY-singlet phase with $\ff S_{1} \ff S_{2} \approx -\frac34$ (see Figs.\ \ref{fig:toyss} and \ref{fig:toykondo}).
In both cases, the ground state is a spin singlet. 
However, the eigenvalue of the $x$ component of the pseudo-spin, the pseudo-charge $q$, is different: 
We have $q=+\frac12$ in the RKKY singlet and $q=-\frac12$ in LKS ground state, while $q=0$ for the PKS state.

Compared to the phase diagram for large $L$, there are two noteworthy points:
(i) There is no RKKY-triplet phase in the $L=3$ model. 
For weak $J$ and small $\Delta$, we rather find a competition between the RKKY-singlet phase, characterized by strong antiferromagnetic impurity-spin correlations $\langle \ff S_{1} \ff S_{2} \rangle$, and the PKS phase.
The impurity-spin correlations in the PKS phase gradually increase from $\langle \ff S_{1} \ff S_{2} \rangle \approx 0$, deep in the PKS regime (strong $J$, large $\Delta$) to {\em ferromagnetic} $\langle \ff S_{1} \ff S_{2} \rangle \to + \frac14$ for $J\to 0$ and small $\Delta$. 
(ii) With decreasing $\Delta$ and down to $\Delta=0$ this PKS phase extends over the entire $J$ range. 
In particular, the phase boundary between the PKS and the RKKY-singlet phase is linear in $J$ for weak $J$, while for $J \to \infty$ and small $\Delta$ the boundary, $\Delta''_{\rm c}(J)$, exhibits a $1/J$ trend. 

We can understand the weak-$J$ regime by means of perturbation theory in $J$, which is regular as the system is gapped, either due to the finite-size gap $\delta>0$ at $\Delta=0$ or due a finite superconducting gap $2\Delta >0$. 
The RKKY approach described in Sec.\ \ref{sec:rkky} equally applies to both, to $\Delta \gg \delta$, i.e., to the case where finite-size effects are irrelevant, and to the case $\delta \gg \Delta > 0$, i.e., to quantum-box physics.
Conceptually, however, the perturbative treatments are different in the two cases $\Delta=0$ and $\Delta>0$. 

For $\Delta>0$, we find that the Kondo effect is absent, as expected, because the $J=0$ state of the electron system is a BCS singlet. 
Furthermore and surprisingly, the RKKY coupling turns out to be antiferromagnetic, $J_{\rm RKKY} > 0$, opposed to the ferromagnetic RKKY coupling found in the thermodynamical limit $L\to \infty$ at weak but finite $\Delta$. 
Again, this is a plausible consequence of the sum rule $\sum_{i'=1}^{L} J_{ii'}=0$, i.e., the trivially strongly ferromagnetic (negative) on-site ($d=0$) RKKY coupling must be compensated by (weaker) antiferromagnetic couplings for $d=1$ and $d=2$.
This explains the absence of an RKKY-triplet phase in the toy model. 

Contrary, for $\Delta=0$ and for the small system with $L=3$, we have the situation of a ``Kondo-vs.-RKKY quantum box'', as discussed in Refs.\ \cite{TKvD99,SGP12,SHPM15}. 
Again, perturbation theory in $J$ is regular, but qualitatively different, since there is a finite ``Kondo'' term in the effective Hamiltonian
\be
H_{\rm eff} = J_{\rm K} \sum_{m} \ff S_{m} \ff s_{\rm F} + J_{\rm RKKY} \ff S_{1} \ff S_{2}
\label{eq:heffbox}
\: ,
\ee
see Ref.\ \cite{SGP12} (supplemental material).
The Kondo coupling $J_{\rm K}$ is positive (antiferromagnetic) and {\em linear} in $J$, while $J_{\rm RKKY} \propto J^{2}$, such that 
the finite-size Kondo effect ``wins'' over the RKKY coupling at sufficiently weak $J$. 
The mechanism for the Kondo coupling $J_{\rm K}$ in the toy model is routed in the $J=0$ electronic structure: 
There are three spin-degenerate conduction-electron eigenstates with one-particle energies $-\sqrt{2}t,0,\sqrt{2}t$. 
At half-filling, the one with lowest (highest) energy is fully occupied (unoccupied), while the delocalized and spin-degenerate ``Fermi'' orbital $\ket{\rm F}$ with zero one-particle energy is filled with either a $\sigma=\, \uparrow$ or a $\sigma=\, \downarrow$ electron. 
Hence, the $J=0$ many-body {\em electronic} ground state is degenerate (opposed to the case $\Delta>0$) and provides a {\em single} Kondo screening channel. 
For $d=2$ both impurity spins in fact couple to the electron spin $\ff s_{\rm F} = \frac12 \sum_{\sigma\sigma'} c^{\dagger}_{F\sigma} \ff \tau_{\sigma\sigma'} c_{F\sigma'}$ of the Fermi orbital $\ket{\rm F}$.
However, only one of the impurity spins can be screened, the other one must remain decoupled.
This explains that the ground state is a total-spin doublet and has the form given by Eq.\ (\ref{eq:pks}), i.e., a PKS state but with the difference that the Kondo screening of the impurity spin $\ff S_{m}$ is not local but achieved by the (same) delocalized Fermi orbital for both, $m=1$ and $m=2$. 
This implies that the according PKS doublet ground state takes the form
\be
\ket \Psi_{+} 
= 
\frac{1}{\sqrt2} 
\ket{\rm occ.} \otimes 
\Big( 
\ket{\rm KS}_{1} \otimes \ket{M}_{2}
+
\ket{M}_{1} \otimes \ket{\rm KS}_{2}
\Big)
\: , 
\label{eq:pksbox}
\ee
with $M=\pm \frac12$, which is simply obtained from Eq.\ (\ref{eq:pks}) by replacing both local-Kondo-singlet states $\ket{\rm LKS}_{m}$ by Kondo singlets $\ket{\rm KS}_{m}$ formed with $\ket{\rm F}$, and the BCS states $\ket{\rm BCS}_{\overline{m}}$ by the occupied Fermi sea $\ket{\rm occ.}$ excluding $\ket{\rm F}$, i.e., for $L=3$, by the lowest one-particle energy eigenstate.

From Eq.\ (\ref{eq:pksbox}) we immediately infer, via a direct calculation, that $\langle \ff S_{1} \ff S_{2} \rangle = +\frac14$ at weak $J$, where the finite-size Kondo effect dominates. 
Note, that this impurity correlation is not due the second-order-in-$J$ RKKY coupling but due to the fact that both impurity spins couple antiferromagnetically, via $J_{\rm K} >0$, to the same Fermi orbital and hence are correlated ferromagnetically. 
This explains the above-mentioned numerical result.

Roughly, the finite-size Kondo effect dominates up to a critical local exchange $J_{\rm c}$, where the bulk Kondo temperature $T_{\rm K} \sim e^{-1/J}$ equals the finite-size gap $\delta$. 
At intermediate $J \gtrsim J_{\rm c}$, the RKKY interaction becomes dominant. 
In fact, this turns out as antiferromagnetic, $J_\mathrm{RKKY} = \frac{3J^2}{32\cdot\sqrt 2}$. 
The RKKY interaction competes with the finite-size Kondo effect for weak $J$ and, for strong $J$, with local-Kondo-singlet formation, which leads to $\langle \ff S_{1} \ff S_{2} \rangle \to 0$. 
It is well known \cite{SGP12}, however, that the RKKY coupling becomes fully effective only for larger systems.
This implies that, with increasing $J$, the impurity correlation decreases but remains positive (ferromagnetic) until it vanishes for $J\to \infty$ (see Fig.\ref{fig:toyss}).

We conclude that the phase boundary in the weak $J$ and small $\Delta$ regime is between a PKS doublet and an RKKY-singlet state.
The PKS state has strong ferromagnetic impurity correlations for weak $J$, where the finite-size Kondo effect dominates over the RKKY coupling. 
With increasing $J$ and $\Delta$, the PKS state smoothly evolves from the Kondo-box form given by Eq.\ (\ref{eq:pksbox}) into a state of the form of Eq.\ (\ref{eq:pks}) with vanishing impurity correlations.
Contrary, the RKKY state at finite $\Delta>0$ for $J\to 0$ features antiferromagnetic impurity correlations. 
This results from the perturbational approach starting from the superconducting ($\Delta >0$) phase as described in Sec.\ \ref{sec:rkky}), where the Kondo term is absent.

\subsection{Thermodynamical limit: extended discussion}  

Now that we have understood the mechanisms at work for the $L=3$ toy model, we can qualitatively discuss the $L\to \infty$ limit.
For $\Delta=0$ and with increasing $L$ and thus with decreasing finite-size gap $\delta \sim 1/L$, the critical coupling $J_{\rm c} \propto -1/ \ln \delta$ shrinks to zero.
This means that the weak-$J$ regime, where the finite-size Kondo effect is active, becomes irrelevant for $L\to \infty$. 
Consequently, $J_{\rm RKKY}$ becomes the dominating energy scale at weak $J$. 
While this is antiferromagnetic for small $L$, it must turn into ferromagnetic as the finite-size gap gets smaller. 
This is the analogous mechanism that has been discussed for the $\Delta$ (rather than the $\delta$) dependence of $J_{\rm RKKY}$ in the context of Fig.\ \ref{fig:jofd}.

In the model for $L=3$, the total-spin doublet (PKS) phase extends over the entire $J$ axis for $\Delta \to 0$. 
We expect the same for $\Delta \to 0$ but for larger $L$ such that finite-size effects are absent at any $\Delta>0$, i.e., in the thermodynamical limit $L\to \infty$ at $\Delta=0$.
This view is consistent with our presented DMRG results down to $\Delta =0.1$:

On the strong-$J$ side of the phase diagram and for small $\Delta$, the impurity-spin correlations $\langle \ff S_{1} \ff S_{2} \rangle \to 0$ and the local Kondo correlation $\langle \ff S_{m} \ff s_{i_{m}} \rangle \to -\frac34$ become identical in both, the doublet PKS and the singlet LKS phase (see Fig.\ \ref{fig:nnnss} for $L=83$). 
The transition line $\Delta_{\rm c}''(J)$ in Fig.\ \ref{fig:nnns}, which separates the $S=0$ LKS from the $S=\frac12$ PKS phase at small $\Delta$, decreases with increasing $J$ and tends to $\Delta=0$ for $J\to \infty$. 
One possibility to approach this point in the phase diagram is via paths such that $\Delta > \Delta_{\rm c}''(J)$, i.e., staying in the LKS phase.
In this case two fully localized Kondo singlets at sites $i_{1}$ and $i_{2}$ are formed and, furthermore, two BCS singlets with supports on the two (for $L\to \infty$) half-infinite chains $i=1,...,i_{1}-1$ and on $i=i_{2}+1, ... ,L$, and, finally, a completely localized pseudo-spin-$\eta_{ix}=\frac12$ state $(| {\rm vac.} \rangle - |\uparrow \downarrow \rangle) / \sqrt2$. 
This means that we have a situation, where both impurity spins are screened separately. 
Another possibility is given by paths with $\Delta < \Delta_{\rm c}''(J)$, i.e., staying in the PKS phase.
Since for $J\to \infty$ the impurity and the Kondo correlation functions become the same in PKS and the LKS phases, the physical situation does not change either, except for the point that the localized state at $i_{1}+1$ is a spin doublet $c^\dagger_{i_{1}+1\sigma} | {\rm vac.} \rangle$ ($\sigma=\uparrow, \downarrow$). 
The state is different from but smoothly connected to the state in the deep PKS limit at large $\Delta$ (with $\Delta'_{\rm c}(J) < \Delta < \Delta_{\rm c}(J)$), where the doublet is formed by a single unscreened {\em impurity} spin. 
With decreasing $J$ and at $\Delta = 0$, we expect that the impurity spins are still separately screened but that the two Kondo clouds extend spatially.

Staying in the PKS phase and decreasing $J$ down to $J=0$, however, there is a smooth crossover from a state with separate Kondo screening to a state with strongly ferromagnetic impurity-spin correlations $\langle \ff S_{1} \ff S_{2} \rangle \to +\frac14$ while $\langle \ff S_{m} \ff s_{i_{m}} \rangle \to 0$. 
In the limit $L\to \infty$ and $\Delta\to 0$, the correlations become the same as $J \to 0$ for both, the spin-doublet PKS and in the triplet RKKY phase, which are located on opposite sides, $\Delta < \Delta_{\rm c}(J)$ and $\Delta > \Delta_{\rm c}(J)$, of the transition line.
Hence, the physical picture for $J \to 0$ and $\Delta \to 0$ is the same in both phases.
As the total spin in the PKS is $S=\frac12$, however, the RKKY triplet must be screened down to a doublet state.
Different from the $J\to \infty$ case, the spin doublet localized between the two impurity spins becomes more and more delocalized with decreasing $J$. 
For $J\to 0$, its character is like in the toy model: the two impurity spins are both coupled antiferromagnetically to the same Fermi orbital and are thus correlated ferromagnetically.
Opposed to the toy model, however, we expect that more than just a single Fermi orbital contributes to the screening in the thermodynamical limit $L\to \infty$.

Concluding, our results for finite $\Delta$ and sufficiently large $L$, such that finite-size artifacts are absent, are consistent with a physical picture at $\Delta =0$, where there is a crossover from individual Kondo screening at strong $J$ to screening of the RKKY triplet at weak $J$.
The delocalized total-spin doublet in the state at finite but small $\Delta$ becomes thermodynamically irrelevant and at $\Delta=0$ just corresponds to the spin doublet that must be realized for a system at half-filling and odd $L$.
Finally, starting from even $L$, one arrives at the same conclusions via an analogous discussion.

\subsection{Self-consistency}
\label{sec:self}

So far we have made the usual assumption that the pairing strength $\Delta_{i} \propto \langle c_{i\downarrow} c_{i\uparrow} \rangle$ is homogeneous, $\Delta_{i} = \Delta$. 
In a BCS-type mean-field theory, there is one-to-one correspondence between the pairing strength and the effective attractive interaction. 
This is convenient because it implies that $\Delta$ can be treated as an external parameter.
On the other hand, as a consequence of the competition between local-Kondo-singlet and BCS-singlet formation, one would expect $\Delta_{i}$ to be site dependent in the vicinity of $i_{1}$ and $i_{2}$, depending on the strength of the local exchange $J$. 
To check the impact of this competition, we replace the pairing term $H_{\rm sc}$ by a Hubbard interaction $H_{U}=U\sum_{i} c^{\dagger}_{i\uparrow} c_{i\uparrow} c^{\dagger}_{i\downarrow} c_{i\downarrow}$ in Eq.\ (\ref{eq:ham}) such that the subsequent anomalous mean-field decoupling yields
$H_{U} \mapsto H_{\rm sc}(\ff \Delta) = \sum_{i} \Delta_{i} (c^{\dagger}_{i\uparrow} c^{\dagger}_{i\downarrow} + \mbox{h.c.})$, 
where $\Delta_{i} = U \langle c_{i\downarrow} c_{i\uparrow} \rangle$ and where we use the short-hand notation $\ff \Delta = \{ \Delta_{i} \}_{i=1,...,L}$.
The site dependence of $\Delta_{i}$ is then obtained by a self-consistent calculation for a fixed attractive Hubbard interaction $U<0$:
Starting from an initial parameter set $\ff \Delta$, the ground state of $H(\ff \Delta)=H_{\rm hop} + H_{\rm ex} + H_{\rm sc}(\ff \Delta)$ is calculated and used to update $\Delta_{i} = U \langle c_{i\downarrow} c_{i\uparrow} \rangle$.
This process is repeated until self-consistency is achieved. 
Using the U(1) gauge degree of freedom as discussed in Sec.\ \ref{sec:mod}, we can assume that all $\Delta_{i}$ are real, if all $\Delta_{i}$ have the same phase.
Indeed, this is found to be the case in the self-consistent solution.

We emphasize that this mean-field approach is tentative, as $\Delta_{i}$ is obtained from the interacting Hamiltonian, including $H_{\rm ex}$, such that one cannot expect that the mean-field solution to represent a stationary point of the total-energy functional.
In addition, for a one-dimensional model with small coordination number, the approach will lead to a strong overestimation of the $i$ dependence. 
Therefore, we expect to obtain only qualitative insights. 
Furthermore, to keep the computational effort manageable, we study the phase diagram of the $L=3$ toy model only. 

Consider the pseudo-spin rotation by $\pi$ around the $y$ axis generated by $\eta_{y}$ and represented by the unitary operator $R=e^{i\pi \eta_{y}}$. 
We have $\eta_{ix} \mapsto R \eta_{ix} R^{\dagger} = - \eta_{ix}$, i.e., $\eta_{ix}$ and $R$ are anticommuting.
With $\Delta_{i} = U \epsilon_{i} \langle \eta_{ix} \rangle$ we get $H(\ff \Delta) \mapsto R H(\ff \Delta) R^{\dagger} = H(-\ff \Delta)$.
This immediately implies that if $\ket {\Psi(\ff \Delta)}$ is a ground state of $H(\ff \Delta)$, then the rotated state $\ket {\Psi(-\ff \Delta)} \equiv R \ket {\Psi(\ff \Delta)}$ is a ground state of $H(-\ff \Delta)$ with the same energy $E(\ff \Delta)$.
Furthermore, if $q$ is the pseudo-charge of $\ket {\Psi(\ff \Delta)}$, then $\ket {\Psi(-\ff \Delta)}$ has pseudo-charge $-q$, since $\eta_{x}\ket {\Psi(-\ff \Delta)} = \eta_{x} R \ket {\Psi(\ff \Delta)} = - R \eta_{x} \ket {\Psi(\ff \Delta)} = -q \ket {\Psi(-\ff \Delta)}$.
Finally, the expectation value of $\eta_{ix}$ in the rotated ground state $\ket {\Psi(-\ff \Delta)}$ is
$\langle \Psi(-\ff \Delta) | \eta_{ix} | \Psi(-\ff \Delta) \rangle = \langle \Psi(\ff \Delta) | R^{\dagger} \eta_{ix} R| \Psi(\ff \Delta) \rangle = - \langle \eta_{ix} \rangle$.
Summarizing we can conclude that with $\ff \Delta$ also $-\ff \Delta$ is a self-consistent solution with the same energy but opposite pseudo-charge, and thus the sign of the pseudo-charge becomes irrelevant when treating $\ff \Delta$ self-consistently.

\begin{figure}[t]
\includegraphics[width=0.98\columnwidth]{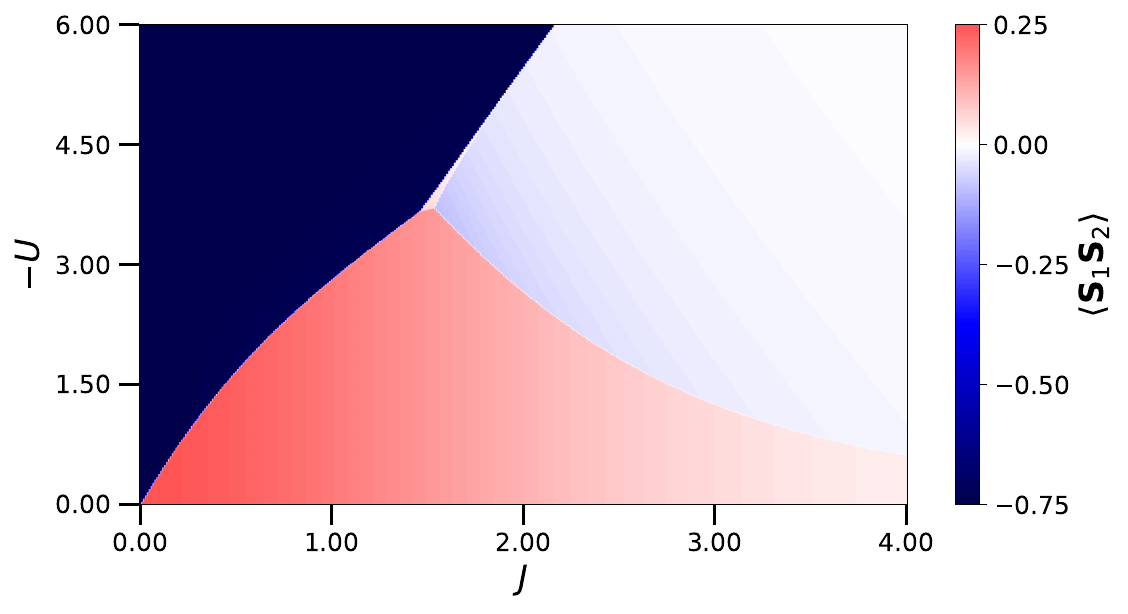}
\caption{
Impurity-spin correlation $\langle \ff S_{1} \ff S_{2} \rangle$ in the 
$(-U)$ vs.\ $J$ phase diagram of the quantum-box model with $L=3$ lattice sites, where $\ff \Delta$ has been calculated self-consistently. 
}
\label{fig:sc}
\end{figure}

Our numerical results are shown in Fig.\ \ref{fig:sc} for comparison with the non-self-consistent results shown in Figs.\ \ref{fig:toy}, \ref{fig:toyss} and \ref{fig:toykondo}.
With Fig.\ \ref{fig:sc} we focus on the impurity-spin correlation $\langle \ff S_{1} \ff S_{2} \rangle$, which is most instructive for the qualitative discussion.
There are four distinct phases with very different impurity-spin correlations.

At weak $J$ and strong $|U|$, we recover the RKKY-singlet phase with strong antiferromagnetic correlations $\langle \ff S_{1} \ff S_{2} \rangle \approx -\frac34$ (blue color). 
The pseudo-charge is $|q|= \frac12$.
The self-consistently computed local pairing strengths $\Delta_{i}$ show a site dependence that reflects the spatial parity symmetry of the $L=3$ model, i.e., $\Delta_1 = \Delta_3 \neq \Delta_2$. 
For strong $|U|$, however, there is not much variation within the RKKY-singlet phase, since all $\Delta_{i}$ approach their maximum value 
$\Delta_{i} / |U| = \langle c_{i\downarrow} c_{i\uparrow} \rangle \to \frac12$. 
The site dependence is only significant in the weak-$U$ limit, where we find $\Delta_{2} \to 0$ as $U\to 0$ while $\Delta_{1}=\Delta_{3}$ remain finite.

At strong $J$ and strong $|U|$, we find the LKS phase with strong antiferromagnetic local Kondo correlations $\langle \ff S_{m} \ff s_{m} \rangle$ for $m=1,2$.
Consequently, the impurity-spin correlation is weak (light blue) and vanishes in the deep LKS limit. 
Hence, the LKS phase has the same spin and pseudo-charge quantum numbers, $S=0$ and $|q|= \frac12$, as the RKKY-singlet phase but very different correlations. 
Again, we have $\Delta_1 = \Delta_3 \neq \Delta_2$, but while $\Delta_{1}=\Delta_{3} \to 0$ for $U,J\to \infty$, we have $\Delta_{2} / |U| \to -\frac12$. 
The suppression of $\Delta_{1}$ and $\Delta_{3}$ is due to the strong local Kondo correlations which enforce fully developed local spin moments $\ff s_{1}^{2}=\ff s_{3}^{2} \to \frac34 \ff 1$ in the LKSs.

There is a small parameter region at intermediate $J$ and strong $|U|$ with weakly ferromagnetic correlations (see pink color in Fig.\ \ref{fig:sc}) and with quantum numbers $S=\frac12$ and $q=0$. 
This is the spin-doublet PKS phase, which, at self-consistency, exhibits a spontaneous breaking of the spatial parity symmetry:
We find $|\Delta_1| < |\Delta_2| < |\Delta_3|$ with a slightly negative $\Delta_{1} < 0$ while $\Delta_2, \Delta_3$ are almost equal and positive. 
In fact, at $i=1$, where the pairing strength $|\Delta_{1}|$ is small, the local Kondo correlation function $\langle \ff S_{1} \ff s_{i_{1}} \rangle$ is strongly antiferromagnetic while $\langle \ff S_{2} \ff s_{i_{2}} \rangle \approx 0$ since $\Delta_{i}$ has the maximum at $i=3$ (note $i_{1}=1$, $i_{2}=3$). 
So this is a PKS state, but compared to Eq.\ (\ref{eq:pks}), the screening is selective and involves only a single impurity spin. 
Note that within the mean-field theory there is a second solution, which can be obtained by mirroring at $i=2$.
In the phase diagram, this total-spin doublet PKS phase separates the RKKY-singlet from the LKS state. 
With increasing $|U|$, the PKS phase becomes narrower in the phase diagram Fig.\ \ref{fig:sc} but its extension remains finite.
So far, the phase diagram thus has the same topology as in the non-self-consistent case discussed in Sec.\ \ref{sec:ed}.

There is, however, one notable difference. 
Namely, in the remaining phase space, at weak $|U|$, there is no stable self-consistent solution with $\Delta_{i} \ne 0$, i.e., the system is in a normal state and, hence, the physics is that of a non-superconducting quantum-box model. 
At $L=3$, the finite-size Kondo effect strongly dominates and produces strong ferromagnetic impurity-spin correlations (red color in Fig.\ \ref{fig:sc}). 
A comparison with the non-self-consistent case is therefore not very meaningful in this parameter range.

In summary, we find the same phases as in the case of a homogeneous paring strength $\Delta$.
The PKS status turns out to be somewhat fragile. 
It undergoes a spontaneous mirror-symmetry breaking and takes up less parameter space in the phase diagram. 

\section{Concluding remarks}
\label{sec:con}

A system of two quantum spins $\frac12$ locally exchange coupled to a conventional s-wave superconductor features a competition between Kondo-singlet formation, BCS-singlet formation, and indirect magnetic exchange, which gives rise to a nontrivial phase diagram.
Since the emergent nonlocal RKKY exchange between the impurity spins involves higher-energy excitations, this competition is expected to depend decisively on the system geometry. 
This dependence cannot be fully captured in approaches where the magnetic exchange is treated within mean-field approaches, such as dynamical mean-field theory \cite{TSRP12}, or where the RKKY coupling is introduced as an independent model parameter \cite{ZLL+10,ZBP11,YMW+14}. 
We have studied the resulting phase diagram for a generic one-dimensional chain geometry, which is accessible to a DMRG approach that fully exploits the non-abelian spin-SU(2) symmetry.

Our results for the one-dimensional model can be summarized as follows:
First, the phase diagrams for impurities coupled to nearest or to next-nearest neighbor sites are fundamentally different. 
In the former case, there are smooth crossovers only with impurity-spin correlations $\langle \ff S_{1} \ff S_{2}\rangle$, ranging from strongly antiferromagnetic for large pairing strengths $\Delta$ to almost vanishing correlations for strong $J$.
Contrary, in the latter case, for an even distance $d=2$, we find discontinuous first-order (level-crossing) transitions between four phases: an RKKY-singlet and a local Kondo-singlet phase with impurity correlations similar as in the $d=1$ phase diagram, but in addition there is a partial Kondo screening (PKS) phase, a total-spin doublet, which separates the two singlet phases in the parameter space, and finally, at weak $J$ and small $\Delta$, an RKKY spin-triplet phase.
As compared to previous NRG calculations \cite{YMW+14}, the topology of the phase diagram is different. 
In particular, the PKS phase is stable in a rather wide range of exchange interactions $J$ at small $\Delta$, and appears to merge with the gapless $\Delta = 0$.

A finite gap, $\Delta>0$, is beneficial for the DMRG calculations because it leads to exponentially decaying RKKY and Kondo correlations at large distances.
In fact, reaching the limit $\Delta \to 0$ is computationally increasingly costly for chains that are sufficiently long to keep finite-size artifacts under control. 
We have thus additionally considered a small system with $L=3$ sites only and two impurity spins at distance $d=2$, which can be treated by exact diagonalization and which gives complementary insights.
Apart from the absence of an RKKY-triplet phase, which is well understood within RKKY perturbation theory, the ground-state phase diagram of this quantum-box model is {\em qualitatively} the same as the DMRG phase diagram for $d=2$, large $L$, and for $\Delta \gtrsim 0.1$. 
In the quantum-box model the $\Delta \to 0$ limit is easily accessible and exhibits a PKS doublet phase, which extends over the whole $J$ range. 
This supports the view that the same holds true for large $L$ in the $\Delta \to 0$ limit and, hence, that the phase boundary between the PKS and the RKKY (singlet or triplet) phase does not end in a critical point at a finite $J$ for $\Delta=0$, i.e., for the gapless system.
We note that, at $\Delta=0$ and for small $L$, a re-entrant finite-size Kondo effect at weak $J$ on a linear-in-$J$ energy scale is present \cite{TKvD99,SGP12}. 
It is known, however, that this is limited to smaller and smaller local exchange couplings $J$ with increasing $L$ and thus eventually becomes irrelevant in the thermodynamical limit $L\to \infty$. 

We have explicitly computed the RKKY exchange coupling $J_{\rm RKKY}$ for the model at finite $\Delta$.
At $L=3$ and $d=2$, for the toy model, this explains the absence of an RKKY-triplet phase for weak $J$. 
More importantly, for large $L$ it explains, in perfect agreement with the full DMRG computations, the critical pairing strength $\Delta_{\rm c}$ that separates the RKKY-triplet phase ($\Delta<\Delta_{\rm c}$) from the RKKY singlet ($\Delta > \Delta_{\rm c}$).
It is remarkable, that the sign of the indirect magnetic exchange can be tuned by varying $\Delta$, since this can be achieved experimentally, e.g., via its temperature dependence.
The distance dependence of $J_{\rm RKKY}$ is interesting as well: 
The well known oscillatory and long-ranged RKKY coupling $\propto (-1)^{d} / d$ of the one-dimensional gapless system at $\Delta=0$ is recovered for $\Delta >0$, but at short distances only.
Beyond a critical $d$ and depending on $\Delta$, the indirect exchange stays antiferromagnetic and, apart from some remaining superimposed oscillations, decays exponentially, i.e., $J_{\rm RKKY} \propto e^{- d/\xi}$ with a length scale $\xi \propto 1/\Delta$.

Using a Lanczos-transformation technique \cite{BMF13,ABMF14,AF19}, it is possible to map impurity-spin systems in rather arbitrary and higher-dimensional lattice geometries onto a one-dimensional chain model.
This opens the perspective for further investigations of similar systems in various geometries, directly relevant to real materials and experimental studies, e.g., of magnetic atoms adsorbed on superconducting surfaces.
In this context, it will be important to additionally consider anisotropic interactions and to extend the studies to multi-orbital models.

Another interesting avenue of further research is to study the effect of the spatial structure of a self-consistently determined pairing strength $\Delta_{i}$. 
Our results for the quantum-box model at $L=3$ indicate that a site-dependent $\Delta_{i}$ does not qualitatively change the phase-diagram topology but affects the parameter space, where the PKS phase is stable. 
In addition, the computation and discussion of the nonlocal pairing correlation function 
$\langle c_{i\downarrow} c_{i\uparrow} \, c_{j\uparrow}^\dagger c_{j\downarrow}^\dagger \rangle$ is worthwhile, especially for the inhomogeneous case with site-dependent $\Delta_{i}$.
Furthermore, a theoretical approach beyond a BCS-type mean-field theory for the conventional superconductor is interesting, see for example Ref.\ \cite{PBZ21}.

\acknowledgments

This work was supported by the Deutsche Forschungsgemeinschaft (DFG, German Research Foundation) through the research unit QUAST, FOR 5249 (project P8), project ID 449872909, and through the Cluster of Excellence ``Advanced Imaging of Matter'' - EXC 2056 - project ID 390715994.

\appendix

\section{Single-impurity model ($M=1$)}
\label{sec:kim}

\begin{figure}[b]
\includegraphics[width=0.85\columnwidth]{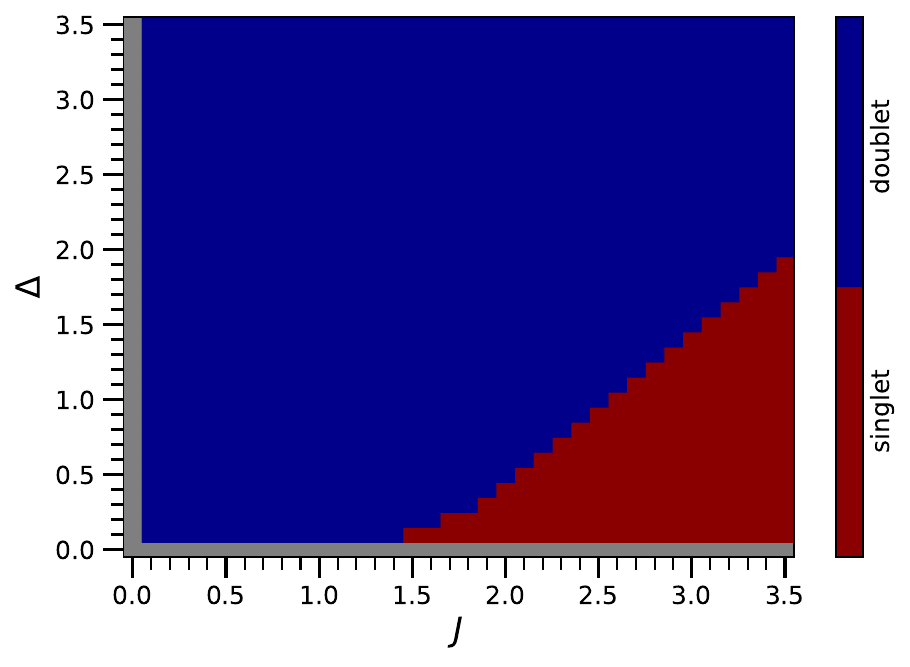}
\caption{
Total-spin quantum number $S$ for a system with $L=41$ lattice sites and a single $s=1/2$ impurity spin coupled to site $i_{1}=21$ of the chain. 
Results as a function of $J$ and $\Delta$.
Red: $S=0$, blue: $S=1$. 
}
\label{fig:kim}
\end{figure}

Figure \ref{fig:kim} shows the DMRG phase diagram for a single impurity spin with $s=\frac12$ exchange coupled to the central site of a chain consisting of $L=41$ sites. 
The Hamiltonian is given by Eq.\ (\ref{eq:ham}), but for $M=1$.
There is a competition between the BCS pairing strength $\Delta$ and the local exchange coupling $J$. 
If $\Delta \gg J$, the electrons of the host system form a BCS-singlet ground state, and the impurity spin remains unscreened. 
Hence, the degenerate ground states 
$\ket \Psi = \ket{\mathrm{BCS}}_{\rm host} \otimes \ket{M = \pm \frac12}_{\rm imp}$
form a total-spin doublet $S=\frac12$, see blue color in the figure.

In the opposite limit $J \gg \Delta$, the impurity spin $\ff S_{1}$ and the local electron spin $\ff s_{i_{1}}$ at the site $i_{1}$ develop strong antiferromagnetic correlations. 
Hence, for $J \to \infty$, there is a local Kondo singlet with $\langle \ff s_{i_{1}} \ff S_{1} \rangle \approx - \frac34$, disentangled from two BCS singlets formed by electrons on the left ($i=1,...,i_{1}-1$) and on the right  ($i=i_{1}+1,...,L$) side of the local Kondo singlet.
The explains that the total ground state is a singlet, $S=0$, see red color in the figure.

The transition line separates two quantum phases with either ``Kondo'' screened ($S=0$) or unscreened impurity spin ($S=\frac12$).
Asymptotically, in the extreme $J\to \infty$ limit, this is given by 
\be
  \Delta_{\rm c} = \frac34 J
  \: ,
\label{eq:cline}  
\ee
as is easily obtained by diagonalization of the Hamiltonian (\ref{eq:ham}) for $M=1$ and $L=1$. 
Across the local quantum phase transition from $S=0$ to $S=\frac12$, there is also a jump of the fermion $\Pi$ parity from odd to even.

\section{BCS Hamiltonian on an open chain} 
\label{sec:bcs}

We consider the following BCS-type Hamiltonian for electrons hopping on a one-dimensional chain with $L$ sites and open boundaries:
\ba
H_{\rm BCS} 
&=& 
-t\sum_{j=1}^{L-1}
\sum_{\alpha=\uparrow,\downarrow}
\left(
c_{j\alpha}^\dagger c_{j+1\alpha} 
+ 
c_{j+1\alpha}^\dagger c_{j\alpha} 
\right) 
\nonumber \\
&+&
\Delta 
\sum_{j=1}^{L}
\left(
c_{j\uparrow}^\dagger c_{j\downarrow}^\dagger + c_{j\downarrow}c_{j\uparrow}
\right) 
\: .
\ea
With Eqs.\ (\ref{eq:hhop}) and (\ref{eq:hsc}), the Hamiltonian reads as $H_{\rm BCS} = H_{\rm hop} + H_{\rm sc}$.
We consider the model at half-filling, i.e., for average total particle number $\langle N \rangle = L$.

With the asymmetric particle-hole transformation 
\be
\begin{pmatrix}
f_{j\uparrow}\\f_{j\downarrow}
\end{pmatrix}
=
\begin{pmatrix}
c_{j\uparrow}\\c_{j\downarrow}^\dagger
\end{pmatrix}
\ee
one can express $H_{\rm BCS}$ in terms of new fermionic operators $f_{j\alpha}, f^{\dagger}_{j\alpha}$ as
\be
H_{\rm BCS} = \sum_{ij\alpha\beta} h_{i\alpha,j\beta} f_{i\alpha}^\dagger f_{j\beta} 
\: , 
\ee
where
\be
h_{i\alpha,j\beta} =\delta_{ij}(1-\delta_{\alpha\beta})\Delta-(\delta_{i,j+1} + \delta_{i+1,j})\delta_{\alpha\beta}\sgn(\alpha)t
\: , 
\ee
and where $\sgn(\uparrow) \equiv +1$, $\sgn(\downarrow) \equiv -1$.

The Hamiltonian can be diagonalized analytically. 
With 
\be
k_{n}
\equiv
\frac{\pi n}{L+1} 
\ee
for $n=1,...,L$, the eigenenergies of the hopping matrix $h$ are given by 
\be
\pm\varepsilon_n=\pm\sqrt{4t^2\cos^2(k_n)+\Delta^2}
\: .
\ee
Both, the positive and the negative eigenenergies are twofold degenerate each, except for odd $L$ at $n=(L+1)/2$, where 
$\pm\varepsilon_n=\pm\Delta$.

The components of the eigenvectors $v^{(n,\pm)}$ of $h$ corresponding to $\pm\varepsilon_n$ are given by 
\begin{align}
v_{i\sigma}^{(n,\pm)} &= \frac{(\delta_{\sigma\uparrow} \pm \delta_{\sigma\downarrow})\sin(ik_n)}{\sqrt{L+1}}\cdot\sqrt{1 \mp \frac{2t\, \mathrm{sgn}(\sigma)\cos(k_n)}{\varepsilon_n}}
\: .
\nonumber \\
\label{eq:states}
\end{align}
Since $H_{\rm BCS}$ commutes with $P \equiv \sum_{i,\alpha} f_{i\alpha}^\dagger f_{L+1-i,\alpha}$ in the one-particle sector, the eigenstates of $h$ are classified by their parity $(-1)^{n+1}$.
Note the relations $v_{i\uparrow}^{(n,+)} = -v_{i\downarrow}^{(n,-)}$ and $v_{i\downarrow}^{(n,+)} = v_{i\uparrow}^{(n,-)}$.

After a second unitary transformation to fermion operators
\be
b_{n,\pm} = \sum_{i,\alpha}v^{(n,\pm)}_{i\alpha} f_{i\alpha} = \sum_i \left(v^{(n,\pm)}_{i\uparrow}c_{i\uparrow} + v^{(n,\pm)}_{i\downarrow}c_{i\downarrow}^\dagger\right) 
\: , 
\ee
the Hamiltonian takes the diagonal form
\be
H_{\rm BCS} = \sum_n \left(\varepsilon_n b_{n,+}^\dagger b_{n,+} - \varepsilon_n b_{n,-}^\dagger b_{n,-}\right)
\: .
\ee
The $b$-operator vacuum, defined as $b_{n,\pm} \ket {\mathrm{vac}_b} = 0$, is related to the $c$-operator vaccum as 
\be
\ket {\mathrm{vac}_b} = c_{1\downarrow}^\dagger \cdots c_{L,\downarrow}^\dagger \ket{\mathrm{vac}_c} 
\: , 
\ee
and hence the ground state of $H_{\rm BCS}$ is given by
\be
\ket 0 = b_{1,-}^\dagger \cdots b_{L,-}^\dagger \ket{\mathrm{vac}_b}
\: .
\label{eq:vac}
\ee
$\ket 0$ is a non-degenerate total-spin singlet state.

\section{Second-order perturbation theory in $J$}
\label{sec:sopt}

The full Hamiltonian of the system is given by $H = H_{\rm BCS} + H_{\rm ex}$, where the local antiferromagnetic ($J>0$) exchange interaction
\be
V \equiv H_{\rm ex} =J \sum_{m=1}^M \boldsymbol S_m \boldsymbol s_{i_m}
\ee 
is treated perturbatively (see, e.g., Ref.\ \cite{EFG+05}). 
Here, $\ff s_{i} = \frac12 \sum_{\alpha\alpha'} c^{\dagger}_{i\alpha} \ff \tau_{\alpha\alpha'} c_{i\alpha'}$ denotes the local spin of the electron system at site $i$, where $\ff \tau$ is the vector of Pauli matrices, and $\ff S_{m}$ is the $m$-th localized quantum-spin $\frac12$ coupling to the electron system at site $i_{m}$.

Trivially, for $M$ localized spins $\ff S_{m}$, the ground-state energy $E_{0}$ of $H$ at $J=0$ is $2^{M}$-fold degenerate. 
We denote the unperturbed ground states by $\ket{0,\boldsymbol\sigma}=\ket{0}\otimes\ket{\boldsymbol\sigma}$, where $\ket{0}$ refers to the non-degenerate electron ground state and $\ket{\ff \sigma} = \ket{\sigma_{1}, ..., \sigma_{m}, ...,  \sigma_{M}}$ to the states of the orthonormal $S_{z}$-standard basis of the system of localized spins.

At first order in $J$ the effective Hamiltonian is given by $H_\mathrm{eff}^{(1)} = P_0 V P_0$, where $P_{0} = \sum_{\boldsymbol\sigma} |0, \ff \sigma\rangle \langle 0, \ff \sigma |$ is the projector onto the ground-state subspace.
Since $|0\rangle$ is a total-spin singlet, we have $\langle 0 | \ff s_{i} | 0 \rangle=0$ and thus $H_\mathrm{eff}^{(1)}=0$.

\begin{widetext}
At second order in $J$, we have 
\be
H_\mathrm{eff}^{(2)} = 
-\sum_{N}^{N\neq 0}\frac{P_0VP_NVP_0}{E_N-E_0}
= - \sum_{\boldsymbol\sigma,\boldsymbol\sigma''} \sum_{\Psi}^{E_\Psi\neq E_0}
\frac{\ket{0,\boldsymbol\sigma}\bracket{0,\boldsymbol\sigma}{V}{\Psi}\bracket{\Psi}{V}{0,\boldsymbol\sigma''}\bra{0,\boldsymbol \sigma''}}{E_{\Psi} - E_0}
\: .
\ee
Here, $E_{N}$ is the $N$-th unperturbed ($J=0$) eigenenergy of $H_{\rm BCS}+V$, and $P_{N}$ denotes the projector onto the corresponding energy eigenspace.
A non-vanishing contribution is only obtained for excited states of the form
$\ket{\Psi} = b_{n,+}^\dagger b_{n',-} \ket{0,\boldsymbol\sigma'}$, 
$\ket{\Psi} = b_{n,+}^\dagger b_{n',+}^\dagger \ket{0,\boldsymbol\sigma'}$, 
and $\ket{\Psi} = b_{n,-} b_{n',-} \ket{0,\boldsymbol\sigma'}$, see Eq.\ (\ref{eq:vac}).
All have the same unperturbed energy $\varepsilon_n + \varepsilon_{n'} + E_0$.
Hence, we have
\begin{align}
H_\mathrm{eff}^{(2)} 
&=-\sum_{\boldsymbol\sigma,\boldsymbol\sigma',\boldsymbol\sigma''}\sum_{n,n'}\Bigg( \frac{\ket{0,\boldsymbol\sigma}\bracket{0,\boldsymbol\sigma}{Vb_{n,+}^\dagger b_{n',-}}{0,\boldsymbol\sigma'}\bracket{0,\boldsymbol\sigma'}{b_{n',-}^\dagger b_{n,+}V}{0,\boldsymbol\sigma''}\bra{0,\boldsymbol \sigma''}}{\varepsilon_n + \varepsilon_{n'} }
\nonumber \\
&\qquad\qquad\qquad+
\frac12 \frac{\ket{0,\boldsymbol\sigma}\bracket{0,\boldsymbol\sigma}{Vb_{n,+}^\dagger b_{n',+}^\dagger}{0,\boldsymbol\sigma'}\bracket{0,\boldsymbol\sigma'}{b_{n',+}b_{n,+}V}{0,\boldsymbol\sigma''}\bra{0,\boldsymbol \sigma''}}{\varepsilon_n + \varepsilon_{n'} }
\nonumber  \\&
\qquad\qquad\qquad+
  \frac 12\frac{\ket{0,\boldsymbol\sigma}\bracket{0,\boldsymbol\sigma}{Vb_{n,-} b_{n',-}}{0,\boldsymbol\sigma'}\bracket{0,\boldsymbol\sigma'}{b_{n',-}^\dagger b_{n,-}^\dagger V}{0,\boldsymbol\sigma''}\bra{0,\boldsymbol \sigma''}}{\varepsilon_n + \varepsilon_{n'} }\Bigg)
\: .  
\end{align}
The factor $1/2$ avoids double counting of terms. 
The matrix elements are readily calculated: 
\begin{align}
\bra{0,\boldsymbol\sigma'}{b_{n',-}^\dagger b_{n,+}V}\ket{0,\boldsymbol\sigma''} &=
\frac J2\sum_{m'}
\Big(v^{(n,+)}_{i_{m'}\uparrow}v^{(n',-)}_{i_{m'}\uparrow}
+v^{(n,+)}_{i_{m'}\downarrow}v^{(n',-)}_{i_{m'}\downarrow}\Big)
\bra{\boldsymbol\sigma'}{S_{m'z}}\ket{\boldsymbol\sigma''}
\: , \nonumber \\
\bra{0,\boldsymbol\sigma'}{b_{n',+} b_{n,+}V}\ket{0,\boldsymbol\sigma''} &=
\frac J2\sum_{m'}
\Big(v^{(n,+)}_{i_{m'}\uparrow}v^{(n',+)}_{i_{m'}\downarrow}
-v^{(n,+)}_{i_{m'}\downarrow}v^{(n',+)}_{i_{m'}\uparrow}\Big)
\bra{\boldsymbol\sigma'}{S_{m'-}}\ket{\boldsymbol\sigma''}
\: , \nonumber \\
\bra{0,\boldsymbol\sigma'}{b_{n',-}^\dagger b_{n,-}^\dagger V}\ket{0,\boldsymbol\sigma''} &=
\frac J2\sum_{m'}
\Big(v^{(n,-)}_{i_{m'}\downarrow}v^{(n',-)}_{i_{m'}\uparrow}
-v^{(n,-)}_{i_{m'}\uparrow}v^{(n',-)}_{i_{m'}\downarrow}\Big)
\bra{\boldsymbol\sigma'}{S_{m'+}}\ket{\boldsymbol\sigma''}
\: .
\end{align}
After eliminating the positive-energy eigenstates by expressing them via their negative-energy counterparts, this yields:
\begin{align}
H_\mathrm{eff}^{(2)} &=
-\sum_{\boldsymbol\sigma,\boldsymbol\sigma',\boldsymbol\sigma''}\sum_{n,n'}\frac{\ket{0,\boldsymbol \sigma}\bra{0,\boldsymbol\sigma''}}{\varepsilon_n + \varepsilon_{n'}}
\sum_{m,m'}\frac{J^2}{4}\left(v_{i_m\uparrow}^{(n,-)}v_{i_m\downarrow}^{(n',-)}-v_{i_m\downarrow}^{(n,-)}v_{i_m\uparrow}^{(n',-)}\right)
\left(v_{i_{m'}\uparrow}^{(n,-)}v_{i_{m'}\downarrow}^{(n',-)}-v_{i_{m'}\downarrow}^{(n,-)}v_{i_{m'}\uparrow}^{(n',-)}\right)
\nonumber \\&\qquad\qquad\times\Bigg(
\bra{\boldsymbol\sigma}{S_{mz}}\ket{\boldsymbol\sigma'}\bra{\boldsymbol\sigma'}{S_{m'z}}\ket{\boldsymbol\sigma''} + \frac 12\bra{\boldsymbol\sigma}{S_{m+}}\ket{\boldsymbol\sigma'}\bra{\boldsymbol\sigma'}{S_{m'-}}\ket{\boldsymbol\sigma''} + \frac 12\bra{\boldsymbol\sigma}{S_{m-}}\ket{\boldsymbol\sigma'}\bra{\boldsymbol\sigma'}{S_{m'+}}\ket{\boldsymbol\sigma''}
\Bigg)
\nonumber\\&=
-\frac{J^2}4
\sum_{\boldsymbol\sigma}\ket{0,\boldsymbol\sigma}\bra{0,\boldsymbol\sigma}
\sum_{m,m'}\sum_{n,n'}\frac{\left(v_{i_m\uparrow}^{(n,-)}v_{i_m\downarrow}^{(n',-)}-v_{i_m\downarrow}^{(n,-)}v_{i_m\uparrow}^{(n',-)}\right)
\left(v_{i_{m'}\uparrow}^{(n,-)}v_{i_{m'}\downarrow}^{(n',-)}-v_{i_{m'}\downarrow}^{(n,-)}v_{i_{m'}\uparrow}^{(n',-)}\right)}{\varepsilon_n + \varepsilon_{n'}}
\nonumber\\&\qquad\qquad\times
\Big(S_{mz}S_{m'z} + \frac 12 S_{m+}S_{m'-} + \frac 12 S_{m-}S_{m'+}\Big)
\sum_{\boldsymbol\sigma''}\ket{0,\boldsymbol\sigma''}\bra{0,\boldsymbol\sigma''}
\nonumber\\&=
P_0\sum_{m,m'} J_{i_mi_{m'}}\boldsymbol S_m\boldsymbol S_{m'} P_0.
\end{align}
We end up with an effective Heisenberg-type (or ``RKKY'') model, where the exchange parameters are given by
\be
J_{i_{m}i_{m'}} = -\frac{J^2}{4}\sum_{n,n'}\frac{\left(v_{i_m\uparrow}^{(n,-)}v_{i_m\downarrow}^{(n',-)}-v_{i_m\downarrow}^{(n,-)}v_{i_m\uparrow}^{(n',-)}\right)
\left(v_{i_{m'}\uparrow}^{(n,-)}v_{i_{m'}\downarrow}^{(n',-)}-v_{i_{m'}\downarrow}^{(n,-)}v_{i_{m'}\uparrow}^{(n',-)}\right)}{\varepsilon_n + \varepsilon_{n'}}
\: .
\label{eq:jmms}
\ee

For a further discussion of the indirect exchange coupling constants (\ref{eq:jmms}), we employ the derived expressions for the one-particle states in Eqs.\ (\ref{eq:states}). 
This yields:
\begin{align}
&v_{i\uparrow}^{(n,-)}v_{i\downarrow}^{(n',-)} - v_{i\downarrow}^{(n,-)}v_{i\uparrow}^{(n',-)}
\nonumber \\& =
\frac{\sin(ik_n)\sin(ik_{n'})}{(L+1)\sqrt{\varepsilon_n\varepsilon_{n'}}}
\left(\sqrt{(\varepsilon_n - 2t\cos(k_n))(\varepsilon_{n'}+2t\cos(k_{n'}))}-\sqrt{(\varepsilon_n+ 2t\cos(k_n))(\varepsilon_{n'}-2t\cos(k_{n'}))}\right)
\end{align}
and therewith Eq.\ (\ref{eq:jrkky}) in the main text:
\begin{align}
&J_{i_{m}i_{m'}}
=
\frac{-J^2}{2(L+1)^2}\sum_{n,n'}\sin(i_mk_n)\sin(i_{m'}k_n)\sin(i_mk_{n'})\sin(i_{m'}k_{n'})\cdot\frac{\varepsilon_n\varepsilon_{n'} - 4t^2\cos(k_n)\cos(k_{n'}) - \Delta^2}{\varepsilon_n\varepsilon_{n'}(\varepsilon_n+\varepsilon_{n'})}
\: .
\label{eq:jrkkycopy}
\end{align}

\end{widetext}

\end{document}